\documentclass[twocolumn]{aastex631}
\usepackage{amsmath}
\usepackage{amssymb}	
\usepackage{CJK}
\usepackage{float}

\newcommand{\diff}{\mathrm{d}}
\newcommand{\wv}{\mathbf{w}}
\newcommand{\wvz}{\mathbf{w}(0)}
\newcommand{\Wv}{\mathbf{W}}
\newcommand{\Pv}{\mathbf{P}}
\newcommand{\pt}{p_{\mathrm{t}}}
\newcommand{\pv}{\mathbf{p}}
\newcommand{\rv}{\mathbf{r}}
\newcommand{\rvb}{\mathbf{r}_{\mathrm{b}}}
\newcommand{\Rv}{\mathbf{R}}
\newcommand{\vv}{\mathbf{v}}
\newcommand{\vvb}{\mathbf{v}_{\mathrm{b}}}
\newcommand{\ubb}{u_{\mathrm{b}}}
\newcommand{\gd}{g_{\mathrm{d}}}
\newcommand{\gk}{g_{\mathrm{k}}}
\newcommand{\gdb}{g_{\mathrm{d,b}}}
\newcommand{\gkb}{g_{\mathrm{k,b}}}
\newcommand{\Ub}{U_{\mathrm{b}}}
\newcommand{\aout}{a_{\mathrm{o}}}
\newcommand{\ain}{a_{\mathrm{i}}}
\newcommand{\pout}{P_{\mathrm{o}}}
\newcommand{\pin}{P_{\mathrm{i}}}
\newcommand{\eout}{e_{\mathrm{o}}}
\newcommand{\ein}{e_{\mathrm{i}}}
\newcommand{\inc}{\theta}
\newcommand{\tkl}{t_{\mathrm{KL}}}
\newcommand{\dth}{\Delta t_{\mathrm{H}}}
\newcommand{\dtb}{\Delta t_{\mathrm{B,k}}}
\newcommand{\ns}{n_{\mathrm s}}
\newcommand{\Recc}{\mathcal{R}_e}
\newcommand{\pert}{\mathcal{R}_{\mathrm {p}}}

\newcommand{\tprof}{t_{\mathrm{w}}}
\newcommand{\Rr}{\mathcal{R}_{\mathrm{r}}}

\begin{document}
\begin{CJK*}{UTF8}{gbsn}
\title{New insight of time-transformed symplectic integrator I: hybrid methods for hierarchical triples}

\author[0000-0001-8713-0366]{Long Wang (王龙)}
\email{wanglong8@sysu.edu.cn}
\affiliation{School of Physics and Astronomy, Sun Yat-sen University, Daxue Road, Zhuhai, 519082, China}
\affiliation{CSST Science Center for the Guangdong-Hong Kong-Macau Greater Bay Area, Zhuhai, 519082, China}



\begin{abstract}

Accurate $N$-body simulations of multiple systems such as binaries and triples are essential for understanding the formation and evolution of interacting binaries and binary mergers, including gravitational wave sources, blue stragglers and X-ray binaries. 
The logarithmic time-transformed explicit symplectic integrator (LogH), also known as algorithmic regularization, is a state-of-the-art method for this purpose.
However, we show that this method is accurate for isolated Kepler orbits because of its ability to trace Keplerian trajectories, but much less accurate for hierarichal triple systems.
The method can lead to an unphysical secular evolution of inner eccentricity in Kozal-Lidov triples, despite a small energy error. 
We demonstrate that hybrid methods, which apply LogH to the inner binary and alternative methods to the outer bodies, are significantly more effective, though not symplectic.
Additionally, we introduce a more efficient hybrid method, BlogH, which eliminates the need for time synchronization and is time symmetric. 
The method is implemented in the few-body code \textsc{sdar}. 
We explore suitable criteria for switching between the LogH and BlogH methods for general triple systems.
These hybrid methods have the potential to enhance the integration performance of hierarchial triples.

\end{abstract}



\section{Introduction}
\label{sec:intro}

Observations have revealed that a significant fraction of stars form within multiple systems, including binaries, triples, and higher-order multiplicities \citep{Sana2012,Duchene2013,Moe2019,Gaia2023,Donada2023,Offner2023}. 
These systems play a crucial role in the long-term dynamical evolution of star clusters \citep[e.g.][]{Henon1961,Heggie1975,Hills1975,Heggie2006}.
Perturbed binaries can evolve into close interacting binaries and mergers, which may serve as progenitors of gravitational wave sources \citep{Downing2010, Banerjee2010, Banerjee2017, Banerjee2018, Banerjee2020, Banerjee2021, Tanikawa2013, Antonini2014, Antonini2017, Bae2014, Rodriguez2016a, Rodriguez2016b, Chatterjee2017, Askar2017, Fujii2017, Park2017, Samsing2018a, Samsing2018b, Samsing2018c, Samsing2024, Kremer2019, DiCarlo2019, Rastello2019, Kumamoto2019, Fernandez2019, Fragione2019a, Antonini2020, Hong2020, Weatherford2021, Wang2021, Tiwari2024, Liu2018, Liu2019, Randall2018}, as well as blue stragglers \citep{Bailyn1995, Lombardi1996, Hurley2001, Sills2005, Perets2009,Perets2012, Leigh2011, Hypki2013, Chatterjee2013, Alessandrini2016, Antonini2016,Fragione2019b} and other exotic objects that generate significant astronomical interest.

Star-by-star numerical $N$-body simulations are commonly employed to study the dynamical evolution of star clusters and the formation of these objects.
The simulations use an integrator to solve the equations of motion for $N$ bodies under Newtonian forces, represented by a $6\times N$ dimensional ordinary differential equation (ODE). 
However, accurately and efficiently integrating the orbits of multiple systems within star clusters presents significant challenges. 
Due to the inverse square law of Newtonian gravity, small time steps are necessary to accurately model the motions of stars as they approach each other. 
This situation is particularly prevalent in highly eccentric binaries, close encounters, and chaotic multiple systems.

As multiple systems can complete a significant number of orbits during their lifetimes in star clusters, ODE solvers must also minimize cumulative integration errors to avoid artificially altering the energy and angular momentum of these systems. 
Such errors can lead to inaccurate predictions of binary evolution.

Moreover, the period distribution of multiple systems spans a wide range from days to Myrs, encompassing serveral magnitudes of timescales \citep[e.g.][]{Kroupa1995a}. 
Tracking the orbital evolution of short-period multiple systems in long-lived massive star clusters, such as globular clusters, is time-consuming for classical ODE solvers.

Several solutions have been developed to address the challenges posed by multiple systems in star cluster simulations. 
The first solution employs a Kepler solver to track the orbital motion of binaries instead of directly solving the ODEs of Newtonian motion. 
This approach simplifies calculations by only requiring the orbital phase to be determined from parameters such as masses, semi-major axes, eccentricities, and times. 
The benefits include reduced computational cost and the elimination of cumulative errors and sigularity issues associated with Newtonian forces. 
\cite{Goncalves2014} introduced a hybrid method using Keplerian-based Hamiltonian splitting and developed the \textsc{sakura} code, which uses a Kepler solver for two-body motions in $N$-body systems. \cite{Hernandez2015} developed a modified algorithm that is symplectic and reversible. 

For multiple systems with one dominant central mass, \cite{Wisdom1991} introduced a symplectic method (Wisdom-Holman method) that separates the Keplerian motion from perturbations by other bodies, allowing the application of a Kepler solver while using a second-order symplectic map for perturbations. 
\cite{Chambers1999} expanded this concept by introducing a hybrid method that accommodates close encounters between massive bodies, leading to the development of the $N$-body code \textsc{mercury}, a state-of-the-art code for simulating the motion of planetary systems. 
Recent works by \cite{Rein2019} and \cite{Hernandez2023} have further refined the switching criteria for this hybrid approach.

The secondary solution involves the use of the secular dynamical method for multiple systems \citep[e.g.][]{Ford2000, Eggleton2001, Fabrycky2007, Hamers2015, Hamers2016, Hamers2018, Hamers2020}. 
This method is particularly efficient for perturbed binaries in hierarchical triple and quadruple systems. 

The third solution directly integrates the motion of multiple systems using a specially designed ODE solver.
One such solver, introduced by \citep{Kustaanheimo1965}, transforms the equations of motion of a binary into a harmonic oscillation form to circumvent sigularity. 
This method is applicable to perturbed binaries, and the state-of-the-art $N$-body code for star clusters, \textsc{nbody6}, uses it \citep{Aarseth2003}.

However, a drawback of these solutions is that they work well for hierarchical systems, where particle orbits can be approximately described by Keplerian dynamics, but they are not easily applicable to chaotic multiple systems.
Another ODE solver capable of handling general multiple systems is the time-transformed symplectic method \citep{Hairer1997}.
Symplectic methods, such as LeapFrog integrators, conserve Hamiltonian during the integration of $N$-body systems.
However, these methods require a constant time step, which can be inefficient for high-eccentric orbits and close encounters, as very small steps are needed to maintain accuracy when two bodies approach periapsis.

To address this, \citep{Hairer1997} introduced the time-transformed symplectic method, allowing the time step to vary according to a time transformation function. This flexibility enables adjustments to time steps according to the separation of two bodies, significantly enhancing the performance of the symplectic method.
\cite{Preto1999} developed an explicit time-transformed symplectic method with a specific time-transformation design, resulting in lower computational costs compared to implicit methods.

Additionally, \cite{Preto1999} and \cite{Mikkola1999} introduced a logarithmic transformation within the explicit method, also known as algorithmic regularization, which can accurately trace the Kepler trajectory with only time phase errors.
\cite{Wang2020c} provided mathematical proof of this feature, indicating that the logarithmic transformation yields a kepler solver.
We will henceforth refer to this method as the LogH method for convenience.

The LogH method has been implemented in several $N$-body codes for multiple systems, including \textsc{archain} \citep{Mikkola1999}, \textsc{sdar} \citep{Wang2020a}, \textsc{mstar} \citep{Rantala2020}, and \textsc{tsunami} \citep{Trani2023}.
The basic LogH method employs a second-order drift-kick-drift map. 
To improve the accuracy of the integrator, the \textsc{sdar} code uses a high-order symplectic method introduced by \cite{yoshida1990} with an optional adaptive step size controller, while the other three codes implement the Bulirsch-Stoer method.
Although the adaptive step size and the Bulirsch-Stoer method may disrupt the symplectic feature, they improve accuracy when evolving certain chaotic multiple systems.
These codes also function as submodules in $N$-body codes for star clusters, such as \textsc{nbody7} \citep{Aarseth2003,Aarseth2012}, \textsc{amuse} \citep{PZ2009}, \textsc{petar} \citep{Wang2020b}, and \textsc{bifrost} \citep{Rantala2023}. 

Despite the success of the LogH method, in this work, we demonstrate that it is inefficient for hierarchical multiple systems due to the breakdown of the Kepler solver feature. 
In fact, accuracy significantly decreases even with a third body that has minimal perturbation on the binary.
This issue has not been addressed in previous implementations of the method, primarily because the Bulirsch-Stoer method guarantees accuracy through iteration to a specified error tolerance.
The low accuracy of the LogH method within the Bulirsch-Stoer framework results in poor performance with many adaptive iterations, masking the problem.
Without a detailed investigation of the iteration process, identifying the low accuracy issue is difficult.
Therefore, it is essential to investigate the time transformation function of the LogH algorithm to enhance its efficiency and accuracy for multiple systems.

In Section~\ref{sec:method}, we present the fundamental mathematical framework of the LogH method. Section~\ref{sec:binary-triple} compares the integration of binary and triple systems, demonstrating a significant loss of accuracy for the latter when using the LogH method. In Section~\ref{sec:hybrid}, we introduce hybrid methods that markedly enhance the accuracy of integrating triple systems. 
Section~\ref{sec:crit} discusses the switching criterion between the LogH and hybrid methods for general multiple systems with one inner binary. 
In Section~\ref{sec:discussion}, we evaluate why the hybrid method is superior and its limitations. Finally, Section~\ref{sec:summary} concludes with a discussion of remaining issues and future improvement plans.

\section{The mathematics of the LogH method} \label{sec:method}

\cite{Preto1999} provides the general formula for the explicit time-transformed symplectic method, which is constructed using the extended Hamiltonian of a Newtonian gravitational $N$-body system. 
The standard Hamiltonian for an $N$-body system is defined as 
\begin{equation}
H(\wv) = T(\pv) + U(\rv),
\label{eq:H}
\end{equation}
where $\wv\equiv(\rv,\pv)$, with $\rv$ and $\pv$ representing coordinates and momenta, and $T$ and $U$ denoting the kinetic and potential energy of all objects, respectively.

The extended Hamiltonian is defined as follows:
\begin{equation}
  \Gamma(\Wv) = g(\Wv)\left[H(\wv,t) - H(\wvz,0)\right],
  \label{eq:gammaorg}
\end{equation}
where $g(\Wv)$ is a time-transformation function and $\Wv$ is the extended phase-space vector.
In this formulation, time $t$ is treated as a coordinate, and the corresponding momentum is $\pt$ initialized to $-H(\wvz, 0)$, representing the initial value of $-H(\wvz,t)$.
We define the extended coordinate vector as $\Rv = (\rv, t)$ and the extended momentum vector as $\Pv = (\rv, \pt)$. 
The kinetic energy then includes $\pt$ as $T(\Pv) = T(\pv) + \pt$.
Thus, $\Gamma(\Wv)$ can be reformulated as 
\begin{equation}
  \Gamma(\Wv) = g(\Wv)\left[T(\Pv) + U(\Rv) \right],
  \label{eq:gamma}
\end{equation}

As $t$ is treated as a coordinate, the equation of motion based on $\Gamma(\Wv)$ is defined as
\begin{equation}
  \frac{\diff \Wv}{\diff s} = \{ \Wv, \Gamma(\Wv) \},
\end{equation}
where $\{\}$ denotes the Poisson bracket, and $s$ is a new differential variable replacing $t$. 

A specific form of $g(\Wv)$ introduced by \cite{Preto1999} is given by
\begin{equation}
  g(\Wv) = \frac{f(T(\Pv)) - f(-U(\Rv))}{T(\Pv) + U(\Rv)} ,
  \label{eq:gorg}
\end{equation}
which results in a separable $\Gamma$:
\begin{equation}
  \Gamma(\Wv) = f(T(\Pv)) - f(-U(\Rv)),
  \label{eq:g}
\end{equation}
allowing the use of explicit symplectic methods such as Leapfrog.

The explicit formulae of the equation of motion are given by 
\begin{equation}
\left \{
  \begin{aligned}
    \frac{\diff \rv }{\diff s} &= f'(T(\pv)+\pt) \frac{\partial T(\pv)}{\partial \pv} \\
    \frac{\diff  t  }{\diff s} &= f'(T(\pv)+\pt)\\
    \frac{\diff \pv }{\diff s} &= f'(-U(\rv,t)) \frac{\partial U(\rv,t)}{\partial \rv} \\
    \frac{\diff \pt }{\diff s} &= f'(-U(\rv,t)) \frac{\partial U(\rv,t)}{\partial t}.
  \end{aligned}
  \right .
  \label{eq:tsi}
\end{equation}

As demonstrated in \cite{Preto1999} and \cite{Mikkola1999}, when $f(x) = \log(x)$ and the drift-kick-drift Leapfrog method with a fixed step $\Delta s$ (sDKD) are applied, the solution accurately follows a Kepler orbit.
In this work, we refer to this as the LogH method.
For isolated multiple systems with conservative potential, $U(\rv,t)$ does not explicitly depend on $t$ and $\pt$ is constant.
The drift and kick steps of the method are described as follows:
\begin{itemize}
    \item Drift:
    \begin{equation}
    \left \{
        \begin{aligned}
            \rv_{k+\frac{1}{2}} & = \rv_{k} + \frac{\Delta s}{2} \left . \gd \right | _{2k} \left .\frac{\partial T(\pv)}{\partial \pv} \right | _{2k} \\
            t_{k+\frac{1}{2}} & = t_{k} + \frac{\Delta s}{2} \left . \gd \right | _{2k} \\
            k & = 0~\mathrm{or}~\frac{1}{2},
        \end{aligned}
    \right .
    \label{eq:drift}
    \end{equation}
    \item Kick:
    \begin{equation}
        \pv_{1} = \pv_{0} + \Delta s \left. \gk \right |_{\frac{1}{2}}\left.\frac{\partial U(\rv)}{\partial \rv} \right |_{\frac{1}{2}},
        \label{eq:kick}
    \end{equation}
\end{itemize}
where the two drift steps are represented by different $k$ values, and
\begin{equation}
    \left \{
    \begin{aligned}
        \gd &= f'(T(\pv)+\pt)  = \frac{1}{T(\pv) + \pt} \\
        \gk &= f'(-U(\rv))  = - \frac{1}{U(\rv)}
    \end{aligned}
    \right .
    \label{eq:gdgk}
\end{equation}
represents the time transformation functions for the drift and kick steps, respectively.
Notice that these functions differ from $g(\Wv)$ in Equation~\ref{eq:gammaorg}.

In the drift step, time $t$ is an integrated variable dependent on $\gd$. 
In a Kepler orbit, $\gd$ is maximized and $\Delta t$ is minimized at periapsis.
Thus, the LogH method is better than the classical drift-kick-drift LeapFrog method with a constant time step $\Delta t$.
From now on, we will refer to this classical method as tDKD, distinguishing it from the sDKD used in the LogH method.

In addition, as shown in \cite{Preto1999} and \cite{Wang2020c}, the LogH method functions as a Kepler solver that advances the eccentric anomaly $E$ while keeping other orbital elements constant. 
Consequently, only the time integration is affected by the second-order approximation (truncation) error of the LeapFrog method, while the solution can exactly trace the Keplerain trajectory while experiencing round-off errors.

The integration step $\Delta s$ has a physical interpretation, as the change in eccentric anomaly ($\Delta E$) \citep{Wang2020b,Wang2020c}:
\begin{equation}
  \Delta s = \mathcal{L} \tan{\frac{\Delta E}{2}},
  \label{eq:ds}
\end{equation}
where $\mathcal{L}$ is the conjugate momenta of the mean anomaly, defined as:
\begin{equation}
  \mathcal{L}  =  \sqrt{\frac{G (m_1 m_2)^2 |a|}{m_1+m_2}},
  \label{eq:L}
\end{equation}
with $a$, $e$, $m_1$, $m_2$ and $G$ representing the semi-major axis, eccentricity, masses of two components, and gravitational constant, respectively.
Thus, $\Delta s$ has units of angular momentum. 
As $\Delta s$ approaches infinity, $\Delta E$ equals $\pi$, indicating a traversal of half an orbit.

\subsection{MA2002 method}

\cite{Mikkola2002} developed a time-transformed LeapFrog method, which is more general than the LogH method, although it may sacrifice symplecticity.
This method defines two time-transformation functions for the drift and kick steps of the LeapFrog approach.
During kick steps, the function is defined as:
\begin{equation}
    \gk^\mathrm{MA} = 1/\Omega (\rv),
\end{equation}
where $\Omega(\rv)$ is an abitrary function of $\rv$.

In drift steps, the function is defined as\footnote{Here, $u$ represents $W$ as defined in Equation~12 of \cite{Mikkola2002}} 
\begin{equation}
    \gd^\mathrm{MA} = 1/u,
    \label{eq:gdu}
\end{equation}
where $u$ is a new integrated variable given by 
\begin{equation}
    \frac{\diff u}{\diff t} = \frac{\partial \Omega(\rv)}{\partial \rv} \cdot \vv,
\end{equation}
with $\vv$ representing the velocity vectors.
Since $u$ depends on $\Omega(\rv)$, it is calculated during the kick steps and used in the drift steps.

For conservative potential, we can set $\Omega (\rv)$ as $-U(\rv)$. 
In this case, the method (MA2002) approximates the LogH method.
Here, $\gk^{\mathrm{MA}}$ is equivalent to the function $\gk$ (Equation~\ref{eq:gdgk}) of the LogH method.
The corrresponding $u$ can be expressed in integral form\footnote{The same definition of $u$ in Equation~40 of \cite{Wang2020b} is missing a negative sign and $\diff t$.}:
\begin{equation}
  u = \int \frac{-\partial U(\rv)}{\partial \rv} \cdot \vv \diff t.
  \label{eq:gt}
\end{equation}
By setting the initial value of $u$ to the initial value of $-U(\rv)$, we have $u \equiv -U(\rv)$.
Due to energy conservation, $T(\pv) + \pt = -U(\rv)$, 
$u$ can be viewed as $T(\pv) + \pt$.
Thus, we can find that 
\begin{equation}
    \frac{1}{u} = \frac{1}{-U(\rv)} = \frac{1}{T(\pv) + \pt},
\end{equation}
which is equivalent to $\gd$ function (Equation~\ref{eq:gdgk}) of the LogH method.
However, in numerical simulations, $T(\pv) + \pt \approx -U(\rv)$, making $u$ an approximation of $T(\pv) + \pt$.

Therefore, the MA2002 method can be viewed as the LogH method with $\gd$ approximated by $\gd^{\mathrm{MA}}$.
The drift and kick steps can be described using the same Equations~\ref{eq:drift} and~\ref{eq:kick}, with $\gk$ defined as in Equation~\ref{eq:gdgk} and $\gd$ replaced by Equation~\ref{eq:gdu}.
The value of $u$ can be calculated during kick steps as:
\begin{equation}
    u_1 = u_0 - \Delta s  \left. \gk  \right |_{\frac{1}{2}} \left.  \frac{\partial U(\rv)}{\partial \rv} \right |_{\frac{1}{2}} \cdot \left ( \frac{\left . \vv \right|_{0} + \left . \vv \right|_{1}}{2} \right ), 
    \label{eq:ukick}
\end{equation}
with $\vv|_1$ calculated in Equation~\ref{eq:kick}.
This formula approximates $u$ assuming that $\partial U(\rv)/\partial \rv$ is constant during one step.

The original LogH method is $\Delta s$ symmetric or time symmetric because reversing $\Delta s$ for backward integration yields the same formulae of integrated quantities (Equations~\ref{eq:drift} and \ref{eq:kick}).
The MA2002 method retains this feature, including the additional integration of $u$ in Equation~\ref{eq:ukick}.
Thus, the MA2002 method also achieves good long-term behavior with respect to the cumulative energy error. 

Due to approximation errors, the computation of $\gd$ in the MA2002 method differs from the exact value in Equation~\ref{eq:gdgk}.
Consequently, the MA2002 method is less accurate than the original LogH method, especially for the Kepler orbits. 
The error differences between the two methods are detailed in the Appendix~\ref{app:error}. 
For isolated binaries, the original LogH method achieves better accuracy, with the $\Gamma$ error mainly caused by round-off errors.
In contrast, the MA2002 solution closely follows the Kepler trajectory, but exhibits some secular error patterns.
However, both methods show similar accuracy for triple systems. 
Thus, the MA2002 method serves as a good approximation to the LogH method and adequately demonstrates its limitations in evolving hierarchical triples in the following sections.

In addition, due to flexible changes in $\Omega(\rv)$ and the definition of $u$ in the MA2002 method, we use the MA2002 method to develop a new algorithm in this work.
To ensure consistent comparisons among methods, the term ``LogH method'' will refer to the MA2002 method in the following content. 

\section{Comparison of the LogH method for binary and triple systems} 
\label{sec:binary-triple}

\begin{figure*}[ht!]
\centering
\includegraphics[width=0.6\linewidth]{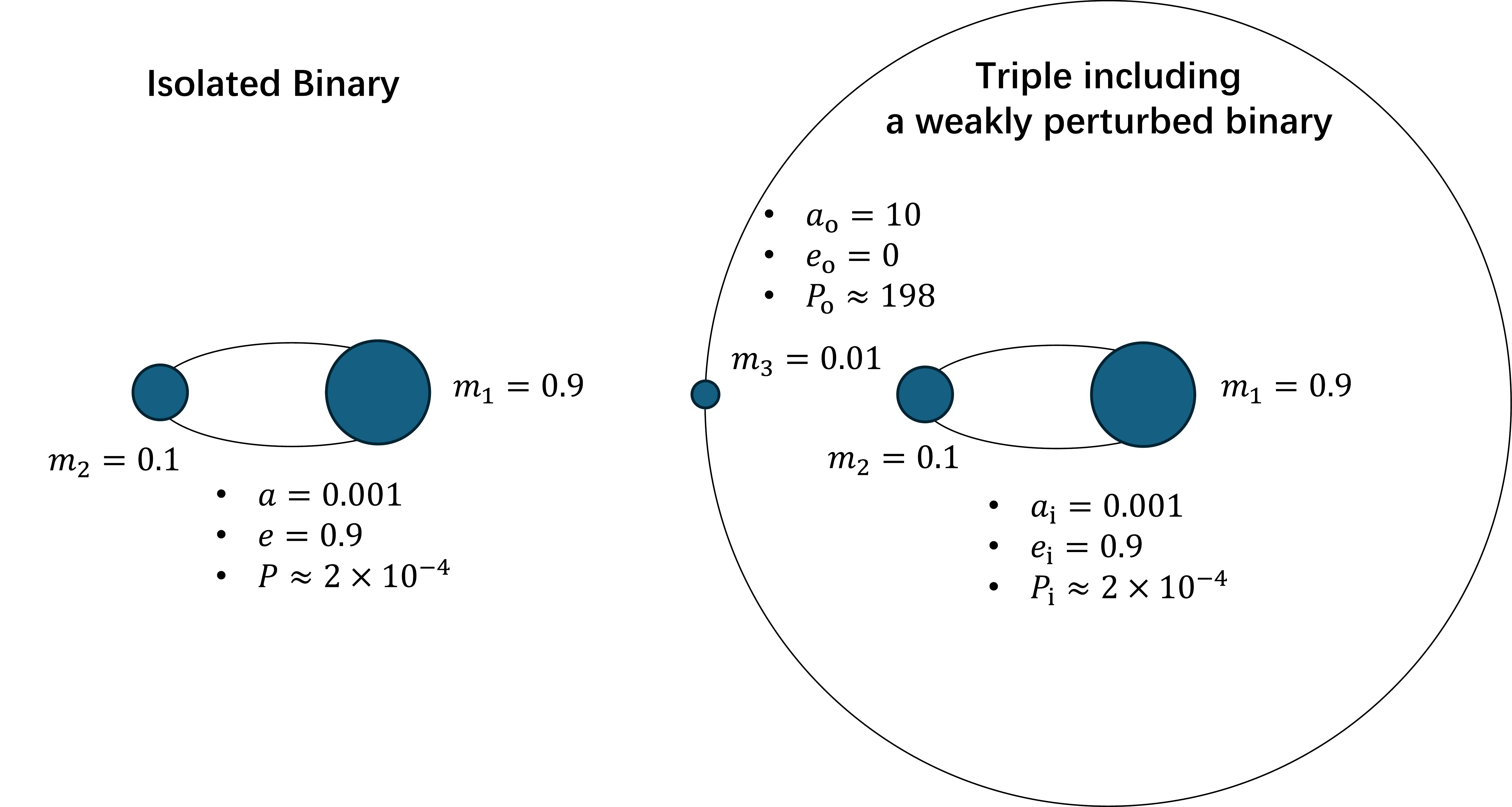}
\caption{The diagram illustrates the initial conditions of the binary and triple for simulations. In the context of the triple system, the subscripts "i" and "O"  in the orbital parameter symbols denote the inner and outer orbits, respectively. Note that there is a fourfold magnitude difference between $\aout$ and $\ain$, and the outer circle and inner ellipse representing the outer and inner orbits of the triple do not reflect the true scale. }
\label{fig:orbitbt}
\end{figure*}

The unique advantage of the LogH method for solving Kepler orbits does not extend to other multiple systems, such as triples. 
Using the \textsc{sdar} code, we compare how the energy error varies with the step size $\Delta s$ for an isolated binary and a hierarchical triple including a weakly perturbed binary. Figure~\ref{fig:orbitbt} illustrates the configurations of the binary and triple systems.

Before detailing the results, we first introduce the model naming convention used throughout the following content. 
To represent values of $\Delta s$, we define a reference value, $\Delta s_0$, which corresponds to one period of the binary. 
We then use the term 'S' followed by the division number to denote the different $\Delta s$ values; for example, S16 represents $\Delta s_0/16$ and S64 represents $\Delta s_0/64$. 
For convenience, we sometimes append this designation as a suffix to the method name to indicate a simulation using that method with the specified $\Delta s$. 
For instance, LogH-S16 denotes the simulation using the LogH method with $\Delta s = \Delta s_0 / 16$. 
To enhance accuracy, we occasionally employ the 6th-order Yoshida method \citep{yoshida1990} in the sDKD and LogH components, indicated as 'Y6' in the model names, such as LogH-Y6-S16. 
This naming convention will be applied consistently throughout the following text.

We set the initial conditions of the binary with an eccentricity of $e=0.9$, a semi-major axis of $a=0.001$ and component masses of $m_1 = 0.9$ and $m_2 = 0.1$.
For convenience, we use scale-free units with gravitational constant $G=1$.
This allows flexibility in scaling the binary to different astrophysical systems by defining two units, such as mass and distance.
The corresponding period of this binary is approximately $2\times10^{-4}$.
The choice of high $e$ and unequal masses better demonstrates the ability of the LogH method to maintain accuracy at periapsis.

\begin{figure}[ht!]
\includegraphics[width=\columnwidth]{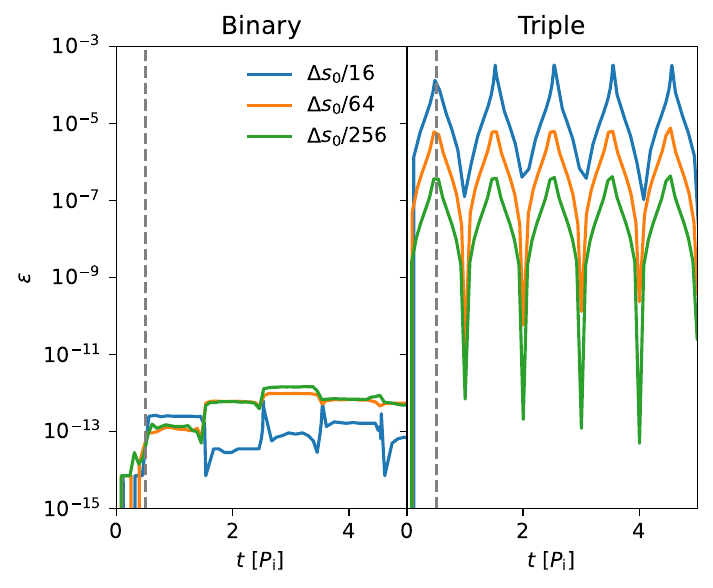}
\caption{The energy error $\epsilon$ as a function of time for simulations of the binary (upper panel) and triple (lower panel) systems using the LogH method, with colors indicating $\Delta s$. The dashed vertical line indicates the time of periapsis. \label{fig:compbintr}}
\end{figure}

Starting at apoaspis, we integrate the binary orbit to $t=0.001$, roughly five periods. 
The cumulative absolute (positive) energy error $\epsilon$ as a function of $t$ is displayed in the left panel of Figure~\ref{fig:compbintr}.
We select three values of $\Delta s$: $1/16, 1/64$ and $1/256$ of $\Delta s_0$ for comparison.
The result shows that $\epsilon$ is on the order of $10^{-13}$ and independent of $\Delta s$, demonstrating that the method traces the Kepler trajectory well.
$\epsilon$ does not exactly reach the round-off error limit of 64-bit double precision. 
The reason is discussed in Appendix~\ref{app:error}.

For comparison, we set the initial conditions of the triple by adding a weak perturber to the binary with the same parameters. 
The perturber is ten times less massive than the lower-mass object in the binary ($m_3 = 0.01$) and orbits the binary in a circular path.
The outer semi-major axis, $\aout = 10$, is $10^4$ times larger than the inner one. 
Both the outer and inner orbits are coplanar with zero inclination.

In this setup, while the influence of the third body is weak, the integration results using the same $\Delta s$ choices as in the isolated binary case exhibit significantly larger $\epsilon$ values depending on $\Delta s$, as shown in the right panel of Figure~\ref{fig:compbintr}. 
The relationship $\epsilon \propto \Delta s^2$ indicates second-order accuracy.
This suggests that even a weak perturber can disrupt the ability to trace Keplerian trajectories, reducing it to a standard second-order symplectic method.

The primary issue when applying the LogH method to triple systems is that the $\gk$ includes two terms of potential energy from the third body:
\begin{equation}
  \gk = f'(-U(\Rv)) = \left [\frac{Gm_1 m_2}{r_{12}} + \frac{Gm_1 m_3}{r_{13}} + \frac{Gm_2 m_3}{r_{23}} \right ]^{-1},
  \label{eq:gktr}
\end{equation}
where $r_{12}$, $r_{23}$ and $r_{13}$ represent the separation between each pair of bodies.
With these terms included, the LogH method cannot be regarded as a Kepler solver like integrator, thus undermining its benefit.

We can view this differently: in the case of an eccentric Kepler orbit, $\gk \propto r_{12}$.
When the binary reaches periapsis, $\gk$ reaches its minimum value, resulting in the smallest $\Delta t$ for a given constant $\Delta s$.
This is reasonable because the acceleration between two bodies is maximized at periapsis, necessitating a small time step for sufficient accuracy.
However, with the presence of a third body, the additional potential terms in $\gk$ smooth the variations of $\gk$, making $\Delta t$ less sensitive to changes in $r_{12}$ of the inner binary.
This can reduce the accuracy of integration at periapsis.

\section{Hybrid methods}
\label{sec:hybrid}

\subsection{H4: Hermite and LogH}

The issue discussed can be avoided in a hybrid scenario. 
In $N$-body codes like \textsc{nbody7}, \textsc{sdar} and \textsc{petar}, the LogH method is combined with other integrators, such as 4th-order Hermite with block time steps, to form a hybrid method referred to as "H4".
In this case, the LogH method is used for the inner binary, while another integrator manages the outer orbit. 
The influence of the third body on the inner binary is treated as a perturbative force during the kick step in the LogH method, allowing the ability to accurately approximate Keplerian trajectories to be preserved.

However, this hybrid method has a disadvantage.
The primary bottleneck is that the time steps $\dtb$ in the LogH method are quantities to integrate, making it difficult to design $\Delta s$ to ensure integration completes at the required time $\dth$ from the other integrator. 
As a result, additional iterative steps are necessary for time synchronization. 
If $\dth$ is comparable to $\dtb$, the hybrid method becomes time-consuming and accumulates timing errors from frequent synchronizations. 
If $\dth$ exceeds the period of the inner binary, the interaction between the inner binary and the outer body occurs less than once per inner orbit, resulting in inaccurate evolution.

\subsection{BlogH: sDKD and LogH}
\label{sec:blogh}
We introduce a new hybrid method "BlogH", which avoids time synchronization.
In this approach, all objects in the triple are integrated using Equations~\ref{eq:drift} and \ref{eq:kick}, but the definitions of $\gd$ and $\gk$ are modified. 
In the BlogH method, only the potential energy of the inner binary,
\begin{equation}
 \Ub = - \frac{ Gm_1m_2}{r_{12}},
\end{equation}
is included in $\gk$ as follows:
\begin{equation}
    \gkb = \frac{1}{-\Ub},
    \label{eq:gkb}
\end{equation}
where new symbol $\gkb$ distinguishes it from $\gk$ in the LogH method.

Since $\gk$ and $\gd$ must represent the equivalent quantities to ensure consistent time steps for kick and drift, and considering total energy conservation, we have
\begin{equation}
    T(\pv) + \pt = -U(\rv,t) = -\Ub + \left[-U(\rv,t) + \Ub \right],
\end{equation}
the corresponding $\gdb$ in the BlogH method is given by:
\begin{equation}
    \gdb  = \frac{1}{T(\pv) + \pt + U(\rv,t)-\Ub},
    \label{eq:gdb}
\end{equation}
which depends on coordinates, making it unsuitable for the explicit method.
The solution is to adopt a similar approach to that in Equation~\ref{eq:gt} to approximate $\gdb$ by defining $\ubb$ as:
\begin{equation}
  \ubb = \int \frac{-\partial \Ub}{\partial \rvb} \cdot \vvb \diff t 
  \label{eq:ublogh}
\end{equation}
where $\vvb$ includes only the inner binary components.
This formula can be evaluated at kick steps following Equation~\ref{eq:ukick}.
Then, 
\begin{equation}
    \gdb = \frac{1}{\ubb},
    \label{eq:gtblogh}
\end{equation}
which can be used in the explicit method.

\begin{figure}[ht!]
\includegraphics[width=\columnwidth]{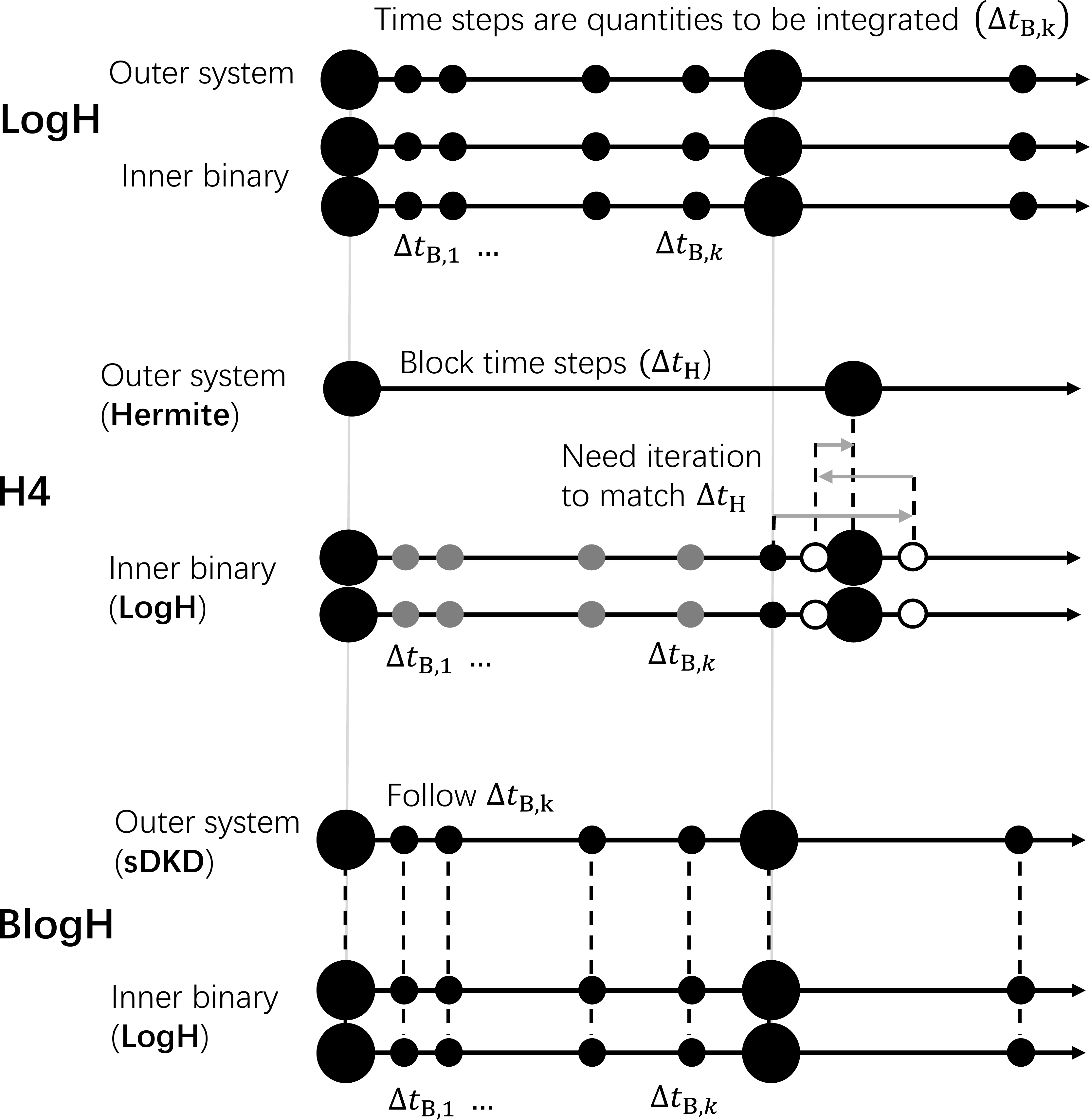}
\caption{The illustration of three integration methods for time step treatment in a multiple system with one inner binary and outer bodies. 
\label{fig:methods}}
\end{figure}

This scenario enables the inner binary to be integrated with the LogH method, while the outer orbit is integrated with the sDKD method, following the same time-step sequences $\Delta t_{\mathrm{B,k}}$ as the inner binary. 
This setup allows the third body to provide perturbations at the kick step without affecting the time transformation.
The method retains the advantages of accurately approximating Keplerian trajectories without the need for time synchronization.
The disadvantage is that the method is not symplectic due to the approximations in the MA2002 method and the unequal time steps of the sDKD loop for outer orbits. 
However, compared to the H4 method, BlogH is time symmetric.
Figure~\ref{fig:methods} illustrates the differences in treatment in time steps among the three methods: LogH, H4 and BlogH.

The BlogH method can also be applied to hierarchical multiple systems that contain one innermost binary.
For example, consider a hierarchical quadruple system with an inner binary with components 1 and 2, and two outer bodies, 3 and 4, which do not form a binary.
The dynamics of the four bodies are integrated using Equations~\ref{eq:drift} and \ref{eq:kick}. However, only the innermost binary (components 1 and 2) is used to calculate $\gkb$ and $\gdb$ through Equations~\ref{eq:gkb}, \ref{eq:ublogh} and \ref{eq:gtblogh}.

Implementing the BlogH method in a LogH $N$-body code is straightforward. 
The only necessary modification is to update the functions $\gd$ and $\gk$, while the overall framework of the integrator remains unchanged.
We have incorporated the BlogH method into the \textsc{sdar} code, alongside the existing LogH and H4 methods. 

\subsection{Triples including a weakly perturbed binary}
\label{sec:triple}

To show that the hybrid methods outperform the LogH method, we redo simulations of the triple system including a weakly perturbed binary by using the LogH, H4 and BLogH meothods with varying $\Delta s$.
We evolve the triple system to 500 periods of the inner binary ($\pin$) to assess the secular evolution of energy and orbital elements.

In the triple system, since the outer orbit is circular, the time step of the Hermite part ($\dth$) can remain constant in the H4 method.
According to the block time step scheme, we select $\dth = 0.5^{14} \approx 0.307 \pin$ for the maximum $\Delta s$.
When $\Delta s$ is reduced by a factor of 4, $\dth$ is halved to maintain consistent accuracy between the second-order LogH method and the fourth-order Hermite method.

\begin{figure}[ht!]
\includegraphics[width=\columnwidth]{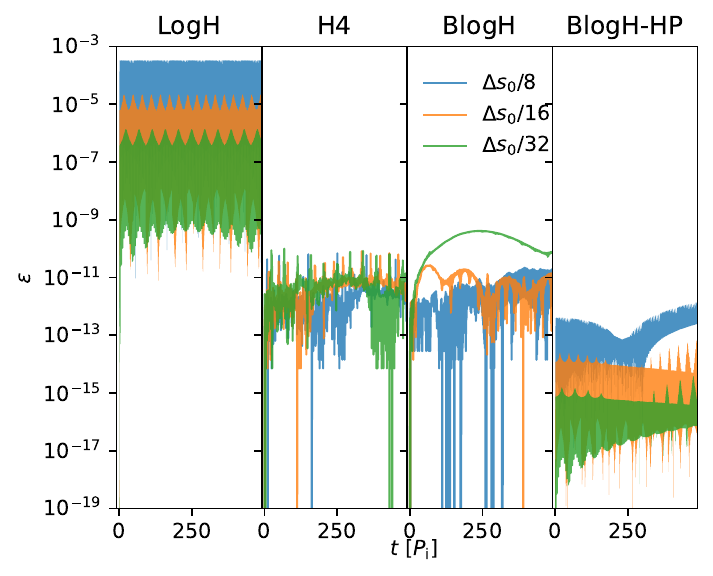}
\caption{The energy errors $\epsilon$ as a function of time are compared for the LogH, H4 BlogH methods, depending on $\Delta s$, in simulating the orbital evolution of the triple system with a weakly perturbed binary. The first three panels display the results for each method, while the last panel features the high-precision BlogH method using the 6th-order Yoshida symplectic method with quadruple precision (30 digits) of floating-point numbers.}
\label{fig:compblogh}
\end{figure}

Comparing the evolution of $\epsilon$ reveals a significant advantage for the H4 and BlogH methods. 
As shown in Figure~\ref{fig:compblogh}, both methods demonstrate much better energy conservation than the LogH method with the same $\Delta s$. 
In particular, $\epsilon$ oscillates significantly in the LogH results, unlike in the other two methods.
Compared to Figure~\ref{fig:compbintr}, the H4 and BlogH methods show a behavior similar to the LogH method for isolated binaries,
suggesting that hybrid methods can effectively approximate the Kepler solution for the inner binaries.

A counterintuitive behavior of the BlogH methods is that a smaller $\Delta s$ leads to a larger $\epsilon$.
This occurs because the influence from round-off errors becomes significant and exceeds the approximation error. 
To investigate this, we performed reference simulations using a high-precision BlogH method (BlogH-HP) which applies quadruple precision with 30 digits. 
As shown in the last panel of Figure~\ref{fig:compblogh}, the same $\Delta s$ results in a lower $\epsilon$, and the relationship between $\Delta s$ and $\epsilon$ aligns with the 2nd-order integration method.

\begin{figure*}[ht!]
\includegraphics[width=\textwidth]{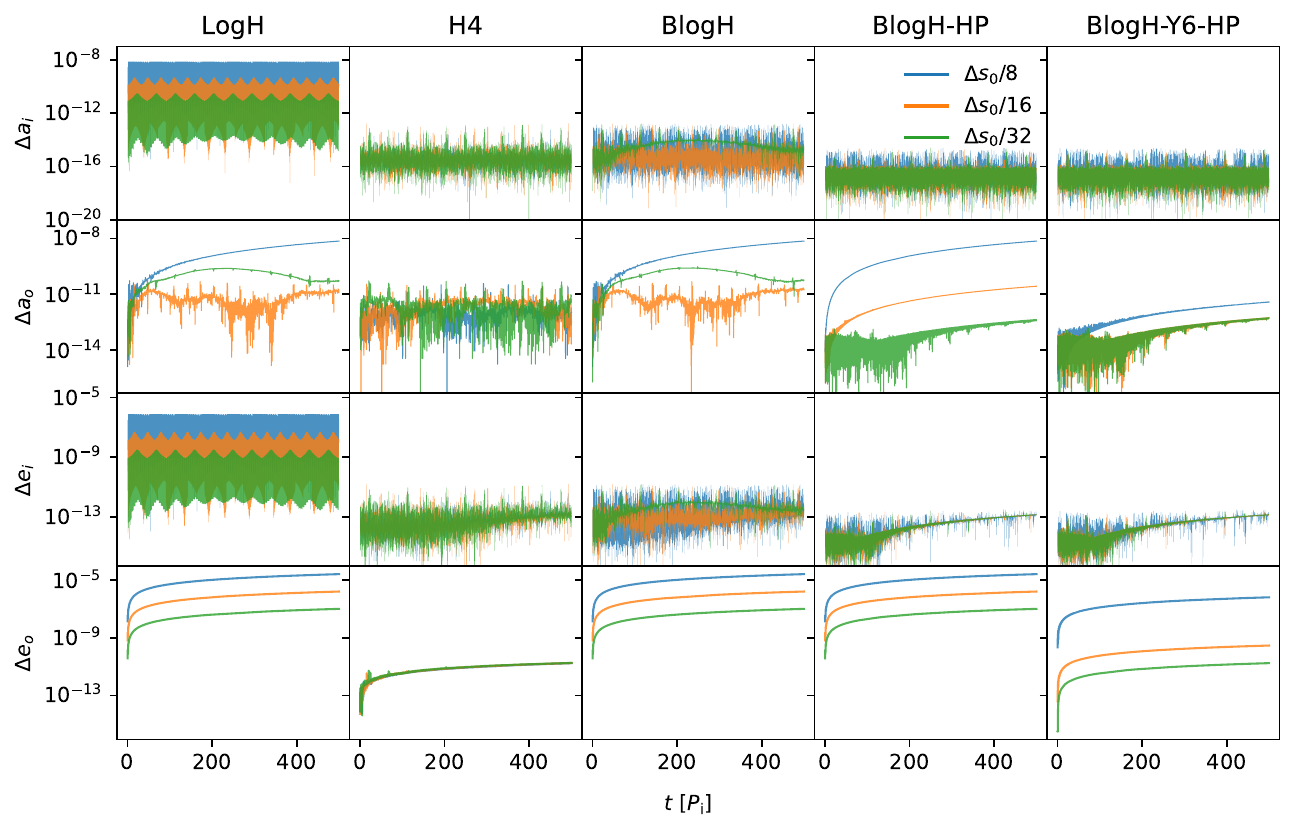}
\caption{The relative evolution of orbital parameters including semi-major axes and eccentricities of the inner and the outer orbits in the triple including a weakly perturbed binary. The columns represent different methods, while the colors indicate various values of $\Delta s$. 
\label{fig:comptrorbit}}
\end{figure*}

For both BlogH and BlogH-HP results, secular patterns emerge in the evolution of $\epsilon$. 
To investigate their origin, we further explore how the evolution of orbital parameters from different methods influences the behavior of  $\epsilon$. 
In the triple system with a weakly perturbed binary, changes in orbital parameters are minimal.
To highlight the differences among the methods in the plot, we define the relative evolution of orbital parameters as 
\begin{equation}
    \Delta o = \left | o - o[0]\right |
    \label{eq:orb}
\end{equation}
where $o$ represents orbital parameters such as $\ain$, $\aout$, $\ein$ and $\eout$, and $o[0]$ denotes the initial value at $t=0$.

The results are presented in Figure~\ref{fig:comptrorbit}. 
In addition to the BlogH-HP models, we introduce the BlogH-Y6-HP models, which use both quadruple precision and the Y6 method to enhance integration accuracy.
The BlogH-Y6-HP-S256 model is the most accurate simulation and serves as a reference. 
By comparing the evolution of $\Delta \ain$ and $\Delta \ein$ across different methods, we observe that the significant oscillation of $\epsilon$ in the LogH models arises from errors in the orbital elements of the inner binary. 
The LogH models exhibit substantial discrepancies compared to the reference model, with smaller $\Delta s$ resulting in reduced discrepancies. 
In contrast, both the H4 and BlogH methods demonstrate comparable inner orbital evolution, aligned closely with the reference model and showing a weak dependence on $\Delta s$.

The evolution of outer orbital elements shows that both LogH and BlogH are similar but significantly larger than the reference model. 
This explains the secular change in $\epsilon$ for the BlogH method (Figure~\ref{fig:compblogh}), which arises from errors in the outer orbital evolution.
Furthermore, for the BlogH-S256 model, the $\epsilon$ curve in Figure~\ref{fig:compblogh} roughly follows the lower boundary of the LogH curve, suggesting that the LogH $\epsilon$ curve combines the oscillations of inner binary errors with the secular component from outer binary errors.

In contrast, the H4 method shows improved outer orbital evolution, with smaller $\Delta \aout$ and $\Delta \eout$ closely aligned with the reference model. 
This indicates that the fourth-order H4 method is more accurate for secular evolution of the outer orbit, even though it is not symplectic.

There is no clear trend regarding how the secular patterns of outer orbital evolution appeared in LogH and BlogH results depend on $\Delta s$. 
One possible explanation is that different $\Delta s$ values lead to varying time phase errors in the inner orbits, resulting in different interactions between the inner and outer orbits and, consequently, different secular patterns.

\begin{table}[ht!]
    \centering
    \caption{The step sizes $\Delta s$ and $\dth$, along with the number of integration steps for the inner binaries in the triple system simulations using the H4 and BlogH methods. The model name suffix indicates the value of $\Delta s$. For the H4 method, both the total steps (tot) and the steps for time synchronization are presented. The LogH method has a similar $\ns$ as the BlogH method and is therefore not included.}
    \begin{tabular}{c|ccc}
        \hline
        Model suffix & S16 & S64 & S256\\
        \hline
        $\Delta s/\Delta s_0$ & $1/16$ & $1/64$ & $1/256$\\
        $\dth/\pin$ & $ 0.307 $ & $ 0.154 $ & $ 0.077 $\\
        $\ns$(H4,tot) & 42824  & 89606  & 227963 \\
        $\ns$(H4,syn) & 22191  & 38424  & 69269 \\
        $\ns$(BlogH) & 8001  & 32004  & 127995 \\
        \hline
    \end{tabular}
    \label{tab:nstep}
\end{table}

Although the BlogH method is less accurate in secular evolution compared to the H4 method, it incurs significantly lower computing costs because of the absence of time synchronization.
We compare the total number of integration steps, $\ns$, reflecting these costs for the H4 and BlogH methods in Table~\ref{tab:nstep}.
The total steps for the H4 method, $\ns(H4,tot)$, are 2-5 times greater that those for the BlogH method, depending on $\Delta s$, 
The steps for time synchronization, $\ns$(H4,syn), account for approximately $1/3-1/2$ of the total steps. 
This suggests that time synchronization significantly affects the performance of the H4 method.

\begin{figure}
    \centering
    \includegraphics[width=\columnwidth]{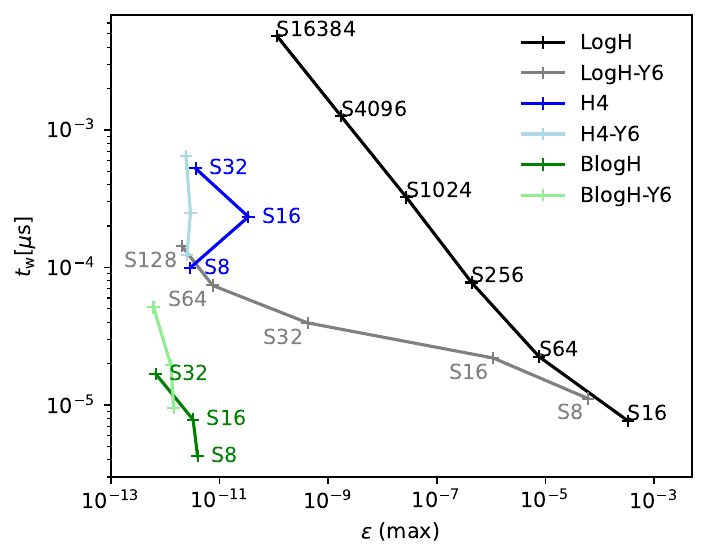}
    \caption{The relationship between maximum $\epsilon$ and wall clock time $\tprof$ for simulations of the triple system, which includes a weakly perturbed binary. The three methods are compared across different $\Delta s$ values (indicated on the plot), both with and without the Y6 method.}
    \label{fig:comptprof}
\end{figure}

The values of $\ns$ can indicate the performance of the algorithm; however, the actual wall clock time ($\tprof$) per inner orbit of the simulation can be influenced by various aspects of implementation.
Therefore, we compare the maximum $\epsilon$ within the evolution of $5~\pin$ versus $\tprof$ for the three methods--LogH, H4 and BlogH--across different $\Delta s$ values, both with and without the Y6 method. 
The results are shown in Figure~\ref{fig:comptprof}.

The LogH results show an increasing trend in \(\tprof\) as both \(\Delta s\) and \(\epsilon\) decrease. 
However, the H4 and BlogH methods have \(\epsilon\) values that are already dominated by round-off errors, so smaller \(\Delta s\) does not significantly reduce \(\epsilon\) and may even lead to an increase.
Therefore, it is preferable to use a larger \(\Delta s\) for the H4 and BlogH methods, such as S8, for better performance.

When only the second-order method is employed, both H4 and BlogH exhibit a few orders of magnitude less \(\tprof\) compared to the LogH method. 
However, when the Y6 method is applied, LogH-Y6 becomes comparable to the H4 method at \(\epsilon \approx 10^{-11}\). 
In contrast, the BlogH method still demonstrates at least one order of magnitude less \(\tprof\), indicating its significant advantage in computational efficiency.

Please note that the absolute values of \(\tprof\) for each method can vary depending on the computing hardware and the implementation of the algorithm. 
The H4 method implemented in the code \textsc{sdar} is significantly more complex than the LogH or BlogH methods; therefore, although the values of \(\ns\) for the H4 method may appear better than those for the LogH method at \(\epsilon \approx 10^{-11}\), the actual \(\tprof\) remains comparable.

\subsection{Kozai-Lidov triples}
\label{sec:kl}

The Hybrid and BLogH methods are accurate not only for triples including a weakly perturbed binary but also significantly enhance accuracy in solving triple systems experiencing Kozai-Lidov oscillations with a highly eccentric inner binary.

When a triple system meets the orbtial criterion for Kozai-Lidov oscillation, the inner eccentricity $\ein$ and the inclination angle $\inc$ between the inner and outer orbits do oscillate on a timescale given by  \cite[e.g.][]{Antognini2015}:
\begin{equation}
  \tkl \approx \frac{8}{15\pi} \left ( 1 + \frac{m_1 + m_2}{m_3} \right ) \frac{\pout^{2}}{\pin} (1 - \eout^{2})^{3/2}.
  \label{eq:tkl}
\end{equation}

\begin{figure}[ht!]
\plotone{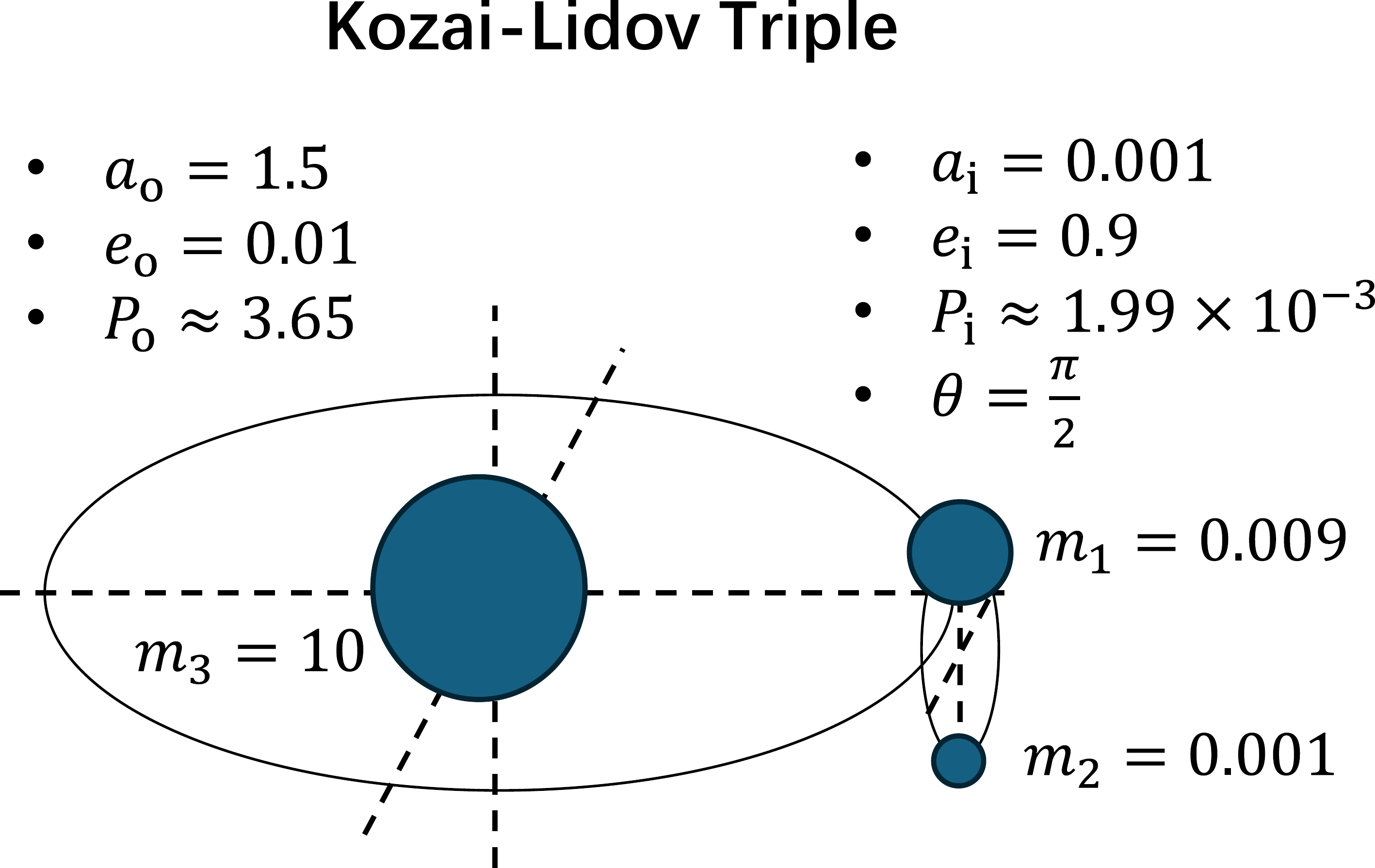}
\caption{The diagram illustrating the initial condition of the triple where the Kozai-Lidov effect induces high eccentricity in the inner binary.
\label{fig:orbitkl}}
\end{figure}

In the following example, we set an initial condition for a Kozai-Lidov triple shown in Figure~\ref{fig:orbitkl}.
The outer body has $m_3 = 10$, significantly more massive than the inner binary with $m_1 = 0.009$ and $m_2 = 0.001$.
The initial inclination $\inc$ is $90^\circ$
and the corresponding $\tkl$ is approximately $1138$.
We can scale the triple to different astrophysical systems.
For instance, with a mass unit of $100~M_\odot$, the triple represents a stellar-mass binary of $0.1$ and $0.9~M_\odot$ orbiting an intermediate-mass black hole of $1000~M_\odot$.
Alternatively, with a mass unit of $0.1~M_\odot$, the triple represents a solar system with binary Jupyter-mass planets.

\begin{figure}[ht!]
\includegraphics[width=\columnwidth]{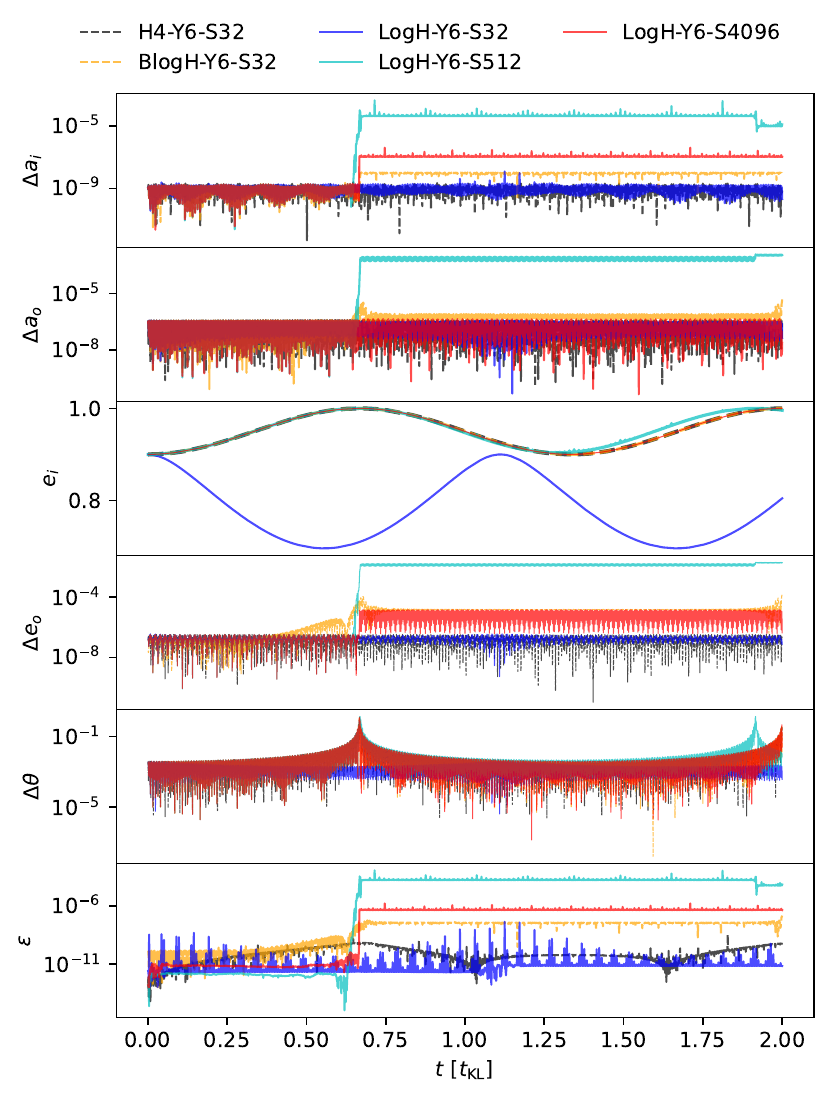}
\caption{The relative evolution of orbital parameters for the Kozai-Lidov triple. The ploting style is similar to that of Figure~\ref{fig:comptrorbit}, with $\ein$ shows as an absolute value instead of $\Delta \ein$. The last row displays $\epsilon$.
\label{fig:compkl}}
\end{figure}

Using the three methods, we integrate the evolution of the triple for about 2 $\tkl$. 
The orbital parameters and $\epsilon$ evolution are shown in Figure~\ref{fig:compkl}.
We compare the results of all the methods with a step size of $\Delta s=\Delta s_0/32$.
The LogH-Y6-S32 model exhibits a markedly different evolution of $\ein$ compared to the H4-Y6-S32 and BlogH-Y6-S32 models, indicating an unphysical result of the LogH method.

To verify this, we add two simulations using the LogH method with much smaller step sizes: S512 and S4096. 
These models show a consistent evolution of $\ein$ with H4-Y6-S32 and BlogH-Y6-S32, confirming that LogH-Y6-S32 presents an unphysical result, despite its lower $\epsilon$.
Moreover, even with smaller steps, the LogH models display jumps in $\ain$, $\eout$ and $\epsilon$ when $\ein$ approaches unity (approximately $0.9999999$), This behavior is less pronounced in the BlogH-Y6-S32 model and absent in the H4-Y6-S32 method, both at the same maximum $\ein$.
This suggests that the LogH method is significantly less accurate than the other two when dealing with high eccentric Kozai-Lidov triples.

\begin{table*}[ht!]
    \centering
    \caption{Step counts ($\ns$) for different methods used to integrate the Kozai-Lidov triple. Both total $\ns$ and $\ns$ per $\pin$ are shown.}
    \begin{tabular}{c|ccccc}
        \hline
        Model name  & BlogH-Y6-S32 & H4-Y6-S32 & LogH-Y6-S32 & LogH-Y6-S4096 & LogH-Y6-S8192\\
        \hline
        $\ns$ & 3.5e+07  & 6.9e+07  & 2.9e+08  & 3.7e+10  & 7.5e+10 \\
        $\ns$[per $\pin$] & 32  & 64  & 269  & 34437  & 68874 \\
        \hline
    \end{tabular}
    \label{tab:klnstep}
\end{table*}

We further compare the step counts of the methods in Table~\ref{tab:klnstep}.
The H4-Y6-S32 model has approximately twice as many as the BlogH-Y6-S32 model, suggesting that time synchronization affects the efficiency of the H4 method, consistent with the findings for triples including a weakly perturbed binary in Section~\ref{sec:triple}.
With the same $\Delta s$, the LogH-Y6-S32 model requires about 7.4 times more steps than the BlogH-Y6-S32 model.
This indicates that the relationship between $\Delta s$ and $\Delta E$ in Equation~\ref{eq:ds} for the Kepler orbit is also disrupted in the LogH-Y6-S32 model.
The LogH-Y6-S8192 model, which exhibits similar accuracy in orbital evolution to the BlogH-Y6-S32 model, needs around 256 times more steps per orbit.
Therefore, the BlogH method demonstrates a significant performance advantage.

\begin{figure}[ht!]
\includegraphics[width=\columnwidth]{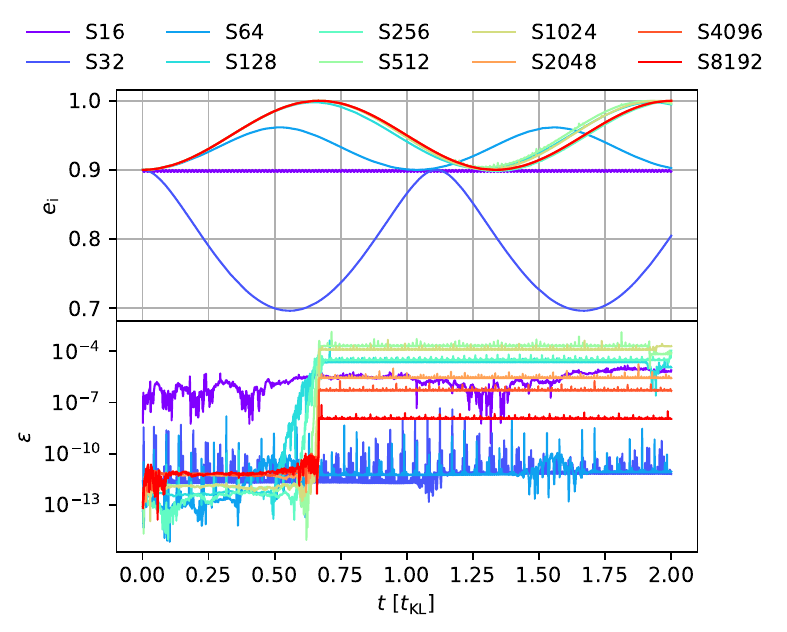}
\caption{The evolution of $\ein$ and $\epsilon$ in the simulations of the Kozai-Lidov triple using the LogH method with varying $\Delta s$.
\label{fig:klein}}
\end{figure}

Two major issues with the low accuracy of the LogH method are the unphysical evolution of $\ein$ and the significant jump in $\epsilon$ when $\ein$ is close to 1.
To investigate how these behaviors depend on $\Delta s$, we perform a series of models using the LogH method with varying $\Delta s$.
The evolutions of $\ein$ and $\epsilon$ are compared in Figure~\ref{fig:klein}.

The S16, S32 and S64 models show significantly different secular evolutions of $\ein$. 
In these simulations, the $\ns$ per $\pin$ are 135, 269 and 538, respectively. 
Despite using the high-accuracy 6th-order Yoshida method, $\ein$ still evolves unphysically.
In addition, repeated peaks appear in the evolution of $\epsilon$ for the S32 and S64 models.
These peaks occur at the periapsis of the inner binary, as illustrated by the LogH results for the triple, including a weakly perturbed binary (Figure \ref{fig:compbintr}).
This suggests that the low accuracy at periapsis from the LogH method introduces an artifical perturbation that repeatedly impacts $\ein$ and leads to an unphysical secular evolution.

Despite the unphysical evolution of $\ein$, $\epsilon$ can still recover to a low value, aligned with the symplectic method.
The maximum $\epsilon$ reaches approximately $10^{-8}$, corresponding to a relative energy error of $10^{-6}$.
Although this relative error may be acceptable for $N$-body simulations of star clusters, it is insufficient for this triple system. 
We also investigated whether the artificial perturbation originates from backward steps in the 6th-order Yoshida method and found no correlation.

As $\Delta s$ increases from S128 to S8192, the evolution trend of $\ein$ converges with the H4 and BlogH results, and jumps in $\epsilon$ begin to appear. 
While the jump value decreases, the average value of $\epsilon$ increases as $\Delta s$ decreases. 
This suggests that a smaller $\Delta s$ can help minimize a large jump but increases cumulative round-off errors.
This inverse relationship between $\Delta s$ and the average $\epsilon$ also occurs in the simulations for the triple including a weakly perturbed binary, as illustrated in Figure~\ref{fig:compblogh}. 

An important feature is that the S32 and S64 models show a similar average $\epsilon$.
Because $\ein$ is far from 1, there is no jump in $\epsilon$. 
In contrast, for the S128 models, the jump occurs at $\ein\approx 1$, resulting in a final $\epsilon$ that is significantly larger than those of the S32 and S64 models.
If we rely solely on the final $\epsilon$ to assess integration accuracy without considering the secular evolution of $\ein$, we may mistakenly conclude that the S32 and S64 models are more accurate.

\begin{figure}
    \centering
    \includegraphics[width=\columnwidth]{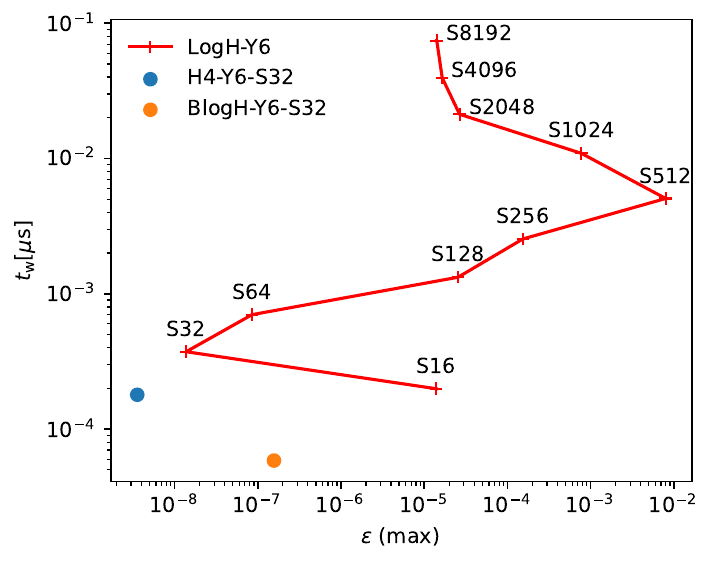}
    \caption{The relationship between maximum $\epsilon$ and wall clock time $\tprof$ for simulations of the Kozai-Lidov triple system. The plotting style is similar to that of Figure~\ref{fig:comptprof}.}
    \label{fig:kztprof}
\end{figure}

We now compare the maximum \(\epsilon\) and \(\tprof\) for the LogH-Y6 models with different \(\Delta s\) values, as well as for the BlogH-Y6-S32 and H4-Y6-S32 models. 
The results are presented in Figure~\ref{fig:kztprof}. 
For the LogH-Y6 method, below S256, the evolution of \(\ein\) is unphysical; consequently, \(\epsilon\) appears low and increases with \(\Delta s\), along with \(\tprof\), except for S16. 
Above S512, energy jumps when \(\ein\) approaches 1 dominate the maximum \(\epsilon\), leading to a more reasonable trend where increasing $\Delta s$ leads to a lower maximum $\epsilon$. 
Comparing this with Figure~\ref{fig:comptprof}, we observe that the H4 and BlogH methods demonstrate greater efficiency at equivalent \(\epsilon\) values when solving the Kozai-Lidov triples with high eccentricities, and the H4 method being notably more efficient than the LogH method.

Note that for the LogH method, the maximum \(\epsilon\) is primarily attributed to the energy jumps. 
It is also possible to reduce the step sizes when \(\ein\) is close to 1 to avoid significant energy jumps, although this breaks the symplecity of the (original) LogH method.
Consequently, with larger \(\Delta s\) values, such as S512, it becomes feasible to obtain a reasonable physical result with much lower maximum $\epsilon$.

\section{Criterion for switching between LogH and BlogH methods}
\label{sec:crit}

The H4 and BlogH methods demonstrate significantly better accuracy than the LogH method for hierarchical triples; however, they lack generality.
As perturbation on the inner binary increases, its motion diverges more from the Kepler orbit, reducing the advantage of the hybrid methods while enhancing the LogH method's performance.
In cases of unstable or chaotic triples where the inner binary cannot be clearly defined, the LogH method is expected to outperform the hybrid methods. 

Since the perturbation on the inner binary depends on $m_3$ and the separation between the third body and the inner binary, we perform a large set of simulations by varying the initial $m_3$ and $\aout$ of based on the initial condition of the Kozai-Lidov triple.
We select 20 values for $m_3$ ranging from $1$ to $10$ and 20 values for $\aout$ from $0.01$ and $0.1$ on a logarithmic scale, generating a total of $20 \times 20$  initial conditions from each pair.
For each initial condition, we perform four simulations using the LogH and BlogH methods with two values of $\Delta s$: S32 and S256. 
The H4 method is excluded due to its similar performance to the BlogH method.

\begin{figure*}[ht!]
\centering
\includegraphics[width=0.8\linewidth]{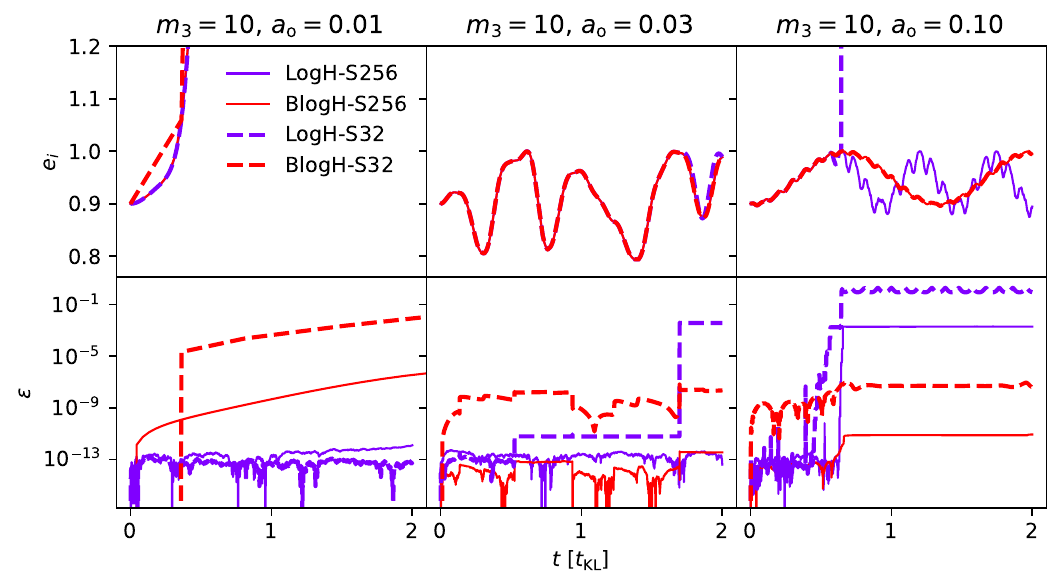}
\caption{The evolution of $\ein$ and $\epsilon$ for three typical simulations of the Kozai-Lidov triple, each with varying $\aout$ values. From left to right, the $\ein$ evolution shows binary disruption, irregular evolution, and regular oscillation. The LogH and BlogH methods are represented in different colors, while different line types indicate the various $\Delta s$. 
\label{fig:klvarysemi}}
\end{figure*}

The secular evolution of $\ein$ in the Kozai-Lidov triple serves as a good indicator to identify the most suitable method.
Figure~\ref{fig:klvarysemi} displays three typical simulations that illustrate the points at which the LogH and BlogH methods change behavior. 
In the left panel, with $m_3 =10$ and $\ain=0.01$, $\ein$ evolve above 1.0, indicating a binary disruption event caused by strong tidal forces from the third object. 
The LogH results with S32 and S256 show a consistent evolution of $\ein$ and produce significantly lower values of $\epsilon$ compared to the BlogH results. 
Therefore, the LogH method is superior to the BlogH method. 

In contrast, the BlogH method outperforms the LogH method in the right panel with $\ain=0.1$. 
The Kozai-Lidov oscillation of $\ein$ with sub-oscillation is observable. 
The LogH-S32 method provides a nonphysical evolution, while the BlogH method exhibits a consistent evolution of $\ein$.
The BlogH-S256 result yields better $\epsilon$ than the LogH-S32 result. 

The middle panel illustrates the transition case, where both methods behave similarly. 
The $\ein$ shows an irregular evolution but remains undisrupted by strong perturbations from the third object. 
The S256 results for both methods exhibit similar $\epsilon$ values.
The LogH-S32 model has a smaller $\epsilon$ than the BlogH-S32 model before $\sim 1.6 \tkl$, but it surpasses it after a jump.

\begin{figure}[ht!]
\includegraphics[width=\columnwidth]{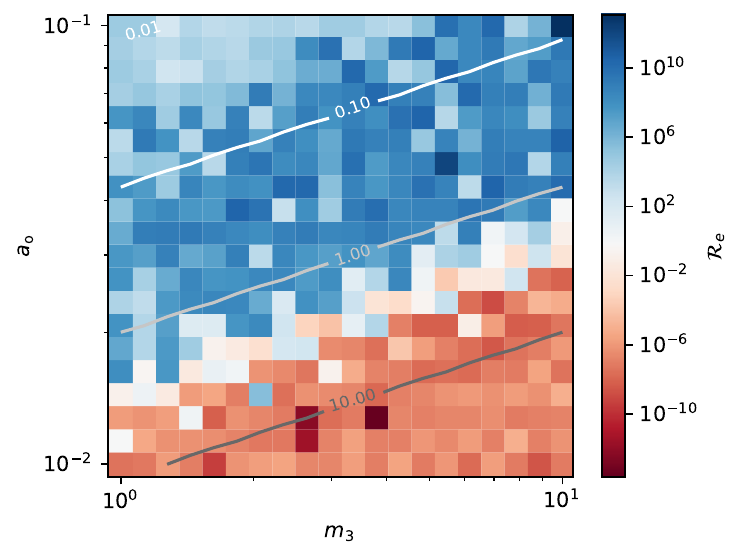}
\caption{The color map displaying the ratio of $\delta \ein$ between the LogH and BlogH results, denoted as $\Recc$, where $\delta \ein$ represents the difference in $\ein$ between S16 and S256 at $2\tkl$. The contour illustrates the perturbation ratio $\pert$ of the outer and inner orbits. 
\label{fig:kleratiomap}}
\end{figure}

To establish the quantitative condition for determining the best method, we define $\delta \ein$ as the absolute difference in $\ein$ at $t=2\tkl$ between the S16 and S256 results. 
A small $\delta \ein$ indicates convergence to a similar final $\ein$, while a large value suggests non-convergence and unphysical behavior. 
We then define the ratio of $\delta \ein$ between the two methods as 
\begin{equation}
    \Recc = \frac{\delta \ein \left [ \mathrm{LogH} \right ]}{\delta \ein \left [ \mathrm{BlogH} \right ]}.
\end{equation}
When $\Recc>1$, the BlogH method shows better convergence in the secular evolution of $\ein$ and is therefore preferred. 
Consequently, the curve where $\Recc = 1$ indicates the switching condition.

Figure~\ref{fig:kleratiomap} displays the color map of $\Recc$, showing a clear gredient related to both $m_3$ and $\aout$.
This gredient trend is well-explained by the tidal perturbation from the third body on the inner binary.
We define the ratio between the outer perturbation and the inner binary's force as follows:
\begin{equation}
    \pert = \frac{m_3 \left(m_1 + m_2\right )}{m_1 m_2}  \left [\frac{ \ain (1 + \ein)}{\aout (1-\eout) }\right ]^3,
    \label{eq:pert}
\end{equation}
which reflects the maximum ratio at the apoapsis of the outer orbit and the periapsis of the inner orbit. 
The contour of $\pert$ in Figure~\ref{fig:kleratiomap} can be described by $\pert = C$, where $C$ is a constant.
The curve of $\Recc=1$ lies within the range of $C$ from 1 to 10.
Thus, we can use the simple switching criterion of $\pert > 1$ to use the LogH method instead of the BlogH for triple systems.

This switching criterion is derived for the Kozal-Lidov triple and may not be suitable for all triple systems.
When the outer orbit is highly eccentric or hyperbolic, the perturbation on the inner binary varies significantly, making this criterion less effective.
We may need a more reliable switching condition based on the separation between the third body and the inner binary.
However, such a criterion could lead to frequent switching as the third body approaches and departs from periapsis.
In future work, a thorough investigation is required to design a more general switching criterion, possibly incorporating the method suggested by \citep{Hernandez2023} to reduce cumulative switching errors. 

\section{Discussion}
\label{sec:discussion}

\subsection{Limitations of the BlogH method}

The absence of time synchronization offers advantages to the BlogH method, but also limits its applicability. 
When two binaries exist in a system, such as a binary-binary quadruple, the LogH method must be applied to each binary with different time transformation functions. 
In this case, the same $\Delta s$ results in different $\Delta t$ values for the two binaries, making time synchronization unavoidable and preventing the use of the BlogH method.
In future studies, we will explore the extension of the BlogH approach to systems with multiple binaries.

\subsection{Combination with the Bulirsch-Stoer method}

To address the low accuracy of the second order sDKD method in the  LogH approach (for both original and MA2002 methods), \cite{Mikkola1999} applied the Bulirsch-Stoer extropolation integrator to create the LogH-BS method.
This method uses polynomial extropolation on a sequence of LogH results obtained through the sDKD loop with varying step sizes of $\Delta s$, predicting an accurate result as the step size approaches zero.

The Bulirsch-Stoer method is efficient when paired with a classic integrator. 
However, the LogH-BS method is less efficient than the pure LogH method for binary systems.
The LogH method can precisely follow the trajectory of a Kepler orbit with only a time-phase error, whereas the extropolated results from the LogH-BS method do not guarantee exact alignment with the trajectory. 
To achieve the same accuracy as the LogH method, the LogH-BS method requires more integration steps. 

The LogH-BS method is more beneficial for systems with $N>2$. 
However, as illustrated in Figure~\ref{fig:klein}, for Kozai-Lidov triples, a low $\epsilon$ does not ensure accurate secular evolution of $\ein$. 
Therefore, only with a strict criterion of $\epsilon <10^{-11}$, can the LogH-BS method potentially provide a physical evolution of $\ein$ while avoiding large errors at the pariapsis of the inner binary.
In contrast, hybrid methods such as the H4 or BlogH method are much more efficient.

The hybrid methods can be combined with the Bulirsch-Stoer method for faster convergence of hierarchical triples.
Further investigation of this possibility is needed in future work.

\subsection{Time Phase Error}

For all three methods, the time phase has errors due to the approximation in time integration, which manifests itself as an artificial phase shift when a large $\Delta s$ is used. 
This error also affects time synchronization in the H4 method. 
Depending on the systems being studied, this error can be significant or negligible. 
In planetary systems where orbital resonances are important, such phase errors may impact long-term evolution and should be carefully investigated. 
However, for multiple systems in star clusters, which is the focus of this work, the error in the time phase is less critical than the errors in energy and angular momentum. 
This is because studies of star clusters primarily concentrate on the statistical evolution of semi-major axes and eccentricities of binaries, which are essential for estimating merger rates and orbital properties of events like gravitational waves. 
Thus, the orbital (time) phase is not important. 
Consequently, orbital average methods that may produce incorrect time phases, such as the slowdown method \citep{Mikkola1996,Wang2020a}, are frequently employed in star cluster simulations to enhance efficiency, including codes like \textsc{nbody6}, \textsc{petar} and \textsc{birfost}. 
More research is needed to explore how orbital average methods can be integrated with the hybrid method.

\section{Summary}
\label{sec:summary}

In this work, we demonstrate that the LogH method (algorithmic regularization) effectively traces Keplerian trajectories in binary systems, but is significantly less accurate for multiple systems like triples. 
As shown in Figure~\ref{fig:compbintr}, even adding a third body with weak perturbation to a binary can introduce significant errors, causing the LogH method to lose the ability to trace Keplerian trajectories and revert to a standard second-order symplectic method. 
Furthermore, the LogH method can lead to an unphysical secular evolution of $\ein$ in a Kozai-Lidov triple, as shown in Figure~\ref{fig:compkl}, unless the step size is extremely small. 
We also found that the energy error, a traditional indicator of integration quality, may fail to identify the unphysical secular evolution of $\ein$ in the Kozal-Lidov triple. 

For hierarchical triples, a more effective solution is to use hybrid methods that combine two integrators, using the LogH method only for the inner binary.
One commonly used hybrid method is H4 (Hermite + LogH), which offers significantly improved accuracy with a step size similar to that of the LogH method; however, it requires time synchronization, adding additional cost.

In this work, we introduce a new hybrid method, BlogH, which combines the LogH method for the inner binary and a sDKD method for outer bodies, using the same time steps.
The BlogH method not only achieves a similar accuracy as H4, but is also more efficient without the need for time synchronization. 
Its implementation is straightforward, requiring only modifications to the time transformation function of the LogH method.
The preliminary implementation of the BlogH method is available in the experimental branch of the \textsc{sdar} code (version 329e).

The main drawback of BlogH is that it is applicable only to multiple systems with one inner binary.
In future work, we will explore a new hybrid algorithm without time synchronization as an extension of the BlogH method for more general $N$-body systems with multiple inner binaries. 

Additionally, the BlogH method becomes less accurate than the LogH method when the inner binary is strongly perturbed.
We can establish a switching criterion to select the optimal method based on the integrated $N$ body system. 
We find that Equation~\ref{eq:pert} serves as a reasonable criterion for this purpose.
In future studies, we will test a wider range of initial conditions for triples to validate and improve the switching criterion.

\section{Acknowledgments}
We thank the very useful comments from the referee. We thank the support of the National Natural Science Foundation of China through grants 21BAA00619 and 12233013, the one-hundred-talent project of Sun Yat-sen University, the Fundamental Research Funds for the Central Universities, Sun Yat-sen University (22hytd09).

%

\vspace{5mm}


\software{numpy (\citealp{harris2020array}),
        matplotlib (\citealp{Hunter:2007}),
        SDAR (\citealp{Wang2020a}, GitHub: https://github.com/lwang-astro/SDAR)
        }



\appendix

\section{Comparison of the original LogH and MA2002 methods}
\label{app:error}

\subsection{Binaries}

Although the original LogH method can trace the Keplerian trajectory with approximation errors affecting only the time phase, as shown by \cite{Preto1999} and \cite{Wang2020c}, $\epsilon$ does not reach the round-off error limit of the floating-point precision.
This occurs because the symplectic map in the LogH method solves the equation of motion using the extended Hamiltonian $\Gamma = \log(T(\Pv)) - \log(-U(\Rv)) = \log(T(\Pv)/-U(\Rv))$. 
Thus, it conserves $\Gamma$ or $T(\Pv)/U(\Rv)$ instead of the original Hamiltonian $H$ or energy, as shown in Equation~43, 44a and 45a of \cite{Preto1999}.
We can approximate $g(\Wv)$ as $f'(-U)$, following Equation 26 in \cite{Preto1999}.
Thus, for the LogH method, we have:
\begin{equation}
    \Gamma \approx  \frac{1}{-U} [H + \pt].
    \label{eq:Gapprox}
\end{equation}
The absolute energy error, $\epsilon = |H + \pt|$, can be approximated by $|-U \Gamma|$.
For an eccentric binary, $-U$ reaches its maximum value at periapsis.
Therefore, although the method ensures that $\Gamma$ is only affected by round-off errors, the error can be significantly amplified by $-U$ when converted to $\epsilon$.
Furthermore, if the separation between the two binary components changes significantly in one step, round-off errors can also significantly affect $\Gamma$, preventing it from reaching the round-off error limit.

\begin{figure}[ht!]
    \centering
    \includegraphics[width=0.7\linewidth]{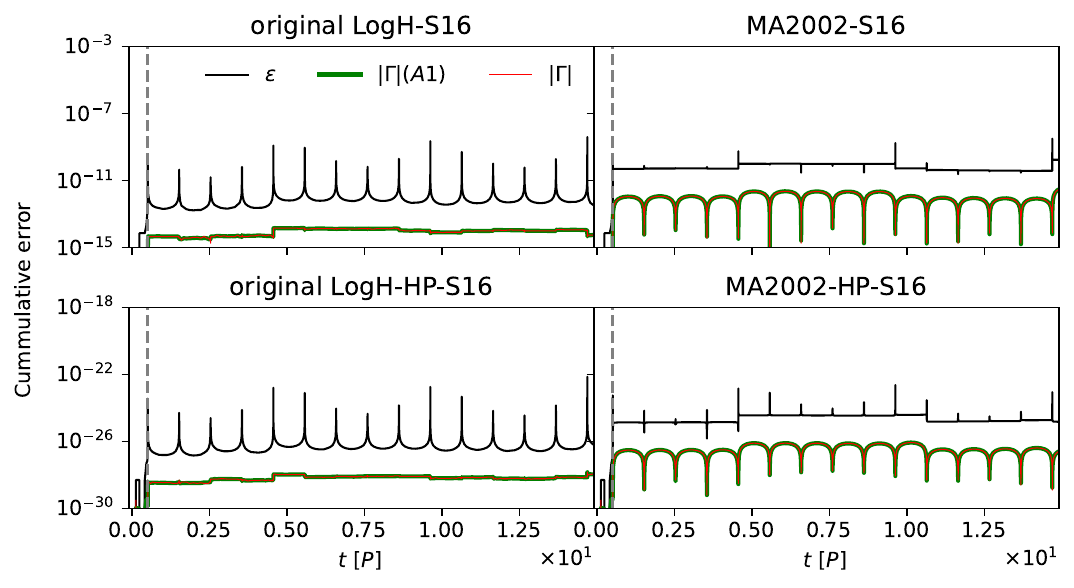}
    \caption{Cumulative absolute integration errors for an extremely eccentric binary ($1-e = 1\times 10^{-8}$) using the original LogH and MA2002 methods. The absolute energy errors ($\epsilon$), the absolute extended Hamiltonian ($|\Gamma|$) and the approximated $|\Gamma|$ from Equation~\ref{eq:Gapprox} are compared. The upper and lower panels show simulations with double precision and 30-dight precision, respectively. The vertical dashed line markes the first periapsis passage.}
    \label{fig:binge}
\end{figure}

To illustrate this behavior, we evolve an extremely eccentric binary with $1-e = 1\times 10^{-8}$ and other initial orbital parameters based on the setup in Figure~\ref{fig:orbitbt}.
Such an extreme case can amplify errors.
The evolutions of $\epsilon$, original $|\Gamma|$, and approximated $|\Gamma|$ (A1) from Equation~\ref{eq:Gapprox} for the original LogH and MA2002 methods are shown in Figure~\ref{fig:binge}.
The upper panels display results with double-precision floating points, while the lower panel shows results with 30-digit precision.
In both cases, \(|\Gamma|\) is two orders of magnitude smaller than \(\epsilon\). 
Based on the orbital parameters, $-U$ is of the order of $10^2$ at apoapsis, consistent with the offset between $\Gamma$ and $\epsilon$.
The results also indicate that \(|\Gamma|\) (A1) nearly overlaps with \(|\Gamma|\), confirming it as a good approximation.

At periapsis, $\epsilon$ exhibits a significant peak due to round-off errors, which is absent in $\Gamma$ for the original LogH method and even decreases for the MA2002 method.
With $-U$ on the order of $10^{10}$ at periapsis, we expect the maximum value of $\epsilon$ to reach $10^{-4}$ at the exact periapsis for double precision results. However, since the simulations do not precisely reach periapsis, the maximum value is significantly lower.

We observe that the MA2002 method is less accurate than the original LogH method due to the approximation of $u$.
This approximation also leads to a distinct pattern in $|\Gamma|$ from periapsis to apoapsis. 
Despite its lower accuracy, the method can still approximate the Keplerian trajectory, significantly outperforming the classical symplectic LeapFrog method for extremely high eccentric binaries.

As the precision increases from double to 30-digit, the magnitude of \(\gamma\) decreases by an order of 14 for both methods, indicating that the error in \(\Gamma\) is mainly due to round-off errors. If the approximation errors were dominant, we would not expect such a significant decrease, as the approximation errors depend mainly on \(\Delta s\), which remains unchanged.

\begin{figure}[ht!]
    \centering
    \includegraphics[width=0.5\linewidth]{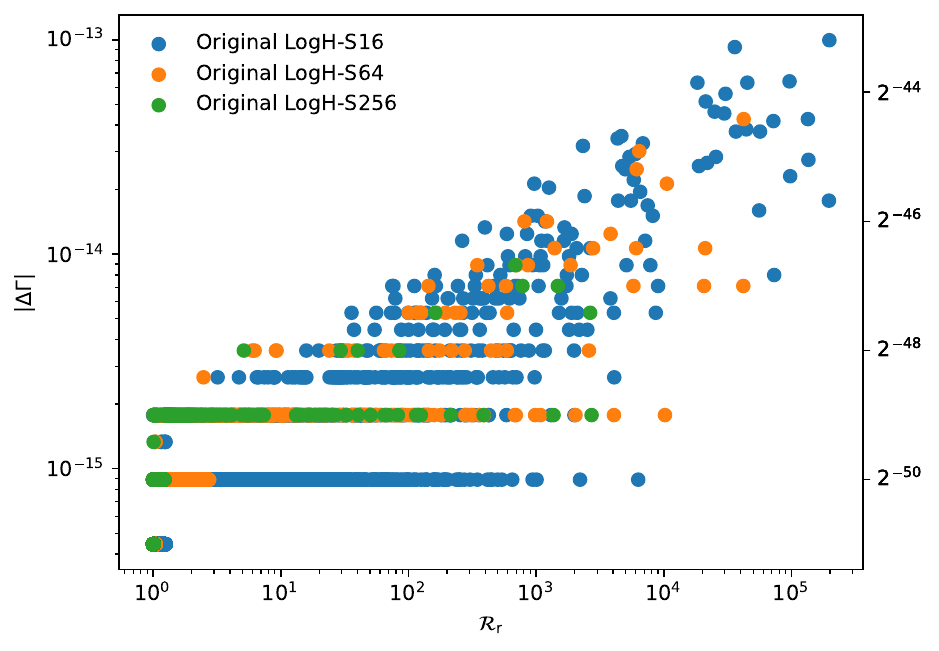}
    \caption{The change in \(\Gamma\) during a single integration step relative to the ratio of the distances between the binary components before and after integration (the ratio of the maximum to the minimum distance). Simulations of the highly eccentric binary using the original LogH methods with varying \(\Delta s\) are compared.}
    \label{fig:dgammaRr}
\end{figure}

In Figure~\ref{fig:binge}, jumps in \(\Gamma\) occur when the orbit approaches periapsis, probably also due to round-off errors. 
As the orbit passes periapsis, the change in distances between the binary components (\(r\)) can be significant. 
According to one drift step of Equation~\ref{eq:drift}, if the new \(r\) is \(10^n\) times smaller than the previous \(r\), the drift calculation introduces \(n\) digits of round-off errors. 
These errors can ultimately affect \(\Gamma\) and may not be recoverable. 

To illustrate this, we investigate how the jumps in $\Gamma$ depend on changes in $r$ for simulations of the extremely high eccentricity binary using the original LogH method with varying \(\Delta s\).
We define \(\Rr\) as the ratio of \(r\) before and after the step (the ratio of the maximum to the minimum \(r\)). 
Figure~\ref{fig:dgammaRr} shows the relationship between $|\Delta \Gamma|$, the change in \(|\Gamma|\) in one step, and \(\Rr\).

The results indicate a correlation between \(|\Delta \Gamma|\) and \(\Rr\).
For a given value of \(\Rr\), \(\Delta \Gamma\) exhibits a wide range of variations, indicating that the correlation between the two is not strict.
This is consistent with the expectation that \(\Delta \Gamma\) arise from round-off errors: Since the rounding errors from changes in \(r\) are random, \(\Delta \Gamma\) does not always experience large jumps, but rather they occur sporadically.

When \(\Delta \Gamma\) is below \(10^{-14}\), the values appear in discrete sequences due to the conversion from binary to decimal. The right \(y\)-axis in Figure~\ref{fig:dgammaRr} displays \(\Delta \Gamma\) values on a logarithmic scale with base 2, where the power index represents individual bits in computer memory. 
This reveals that these discrete sequences correspond to each bit. 
For double precision, the minimum representable value is \(2^{-52}\), which corresponds to approximately \(2.22 \times 10^{-16}\). 
This suggests that the discrete \(\Delta \Gamma\) values reflect the limitations of double precision, indicating that they are influenced by the last few bits of precision. 

The correlation (upper boundary) between \(|\Delta \Gamma|\) and \(\Rr\) is independent of \(\Delta s\), indicating that \(|\Delta \Gamma|\) is unlikely due to approximation errors related to \(\Delta s\) and is instead directly related to \(\Rr\). However, when \(\Delta s\) is small, the likelihood of achieving a large \(\Rr\) decreases, leading to a smaller maximum \(|\Delta \Gamma|\). This suggests that reducing \(\Delta s\) may help minimize maximum \(\Gamma\) errors, but this effect is only seen in the original LogH method. 
In the following analysis, the MA2002 method shows an opposite trend.

\begin{figure}[ht!]
    \centering
    \includegraphics[width=0.7\linewidth]{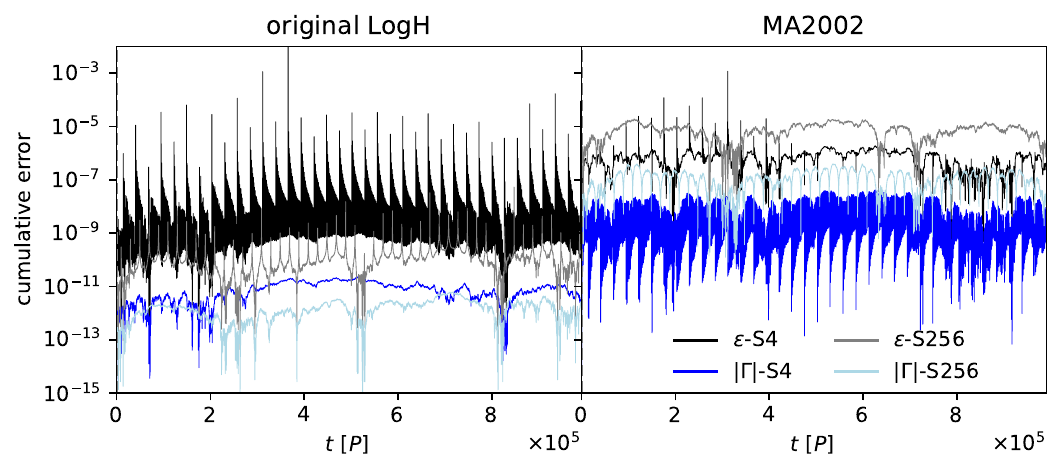}
    \caption{Cumulative absolute integration errors for the extremely eccentric binary ($1-e = 1\times 10^{-8}$) are compared using the original LogH and MA2002 methods, with time integrated over $10^6$ periods and two values of $\Delta s$, S4 and S256.}
    \label{fig:binlong}
\end{figure}

To investigate the long-term behavior of the two methods, we continue to integrate the binary over $10^6$ periods, the cumulative errors are shown in Figure~\ref{fig:binlong}.
For the original LogH method, the two values of $\Delta s$, S4 and S256, result in similar level of $\Gamma$ and $\epsilon$.
The maximum value of errors are lower with S256, due to less maximum $\Rr$.

In contrast, the MA2002 method exhibits an interesting behavior: smaller \(\Delta s\) leads to larger errors. 
This may be due to smaller \(\Delta s\) introducing more cumulative round-off and approximation errors, causing the results to deviate further from the original method.

Although the MA2002 method is not symplectic, it demonstrates time-symmetric behavior during long-term evolution, as there is no significant drift in errors over \(10^6\) periods.

\subsection{Triples}

\begin{figure}[ht!]
    \centering
    \includegraphics[width=0.7\linewidth]{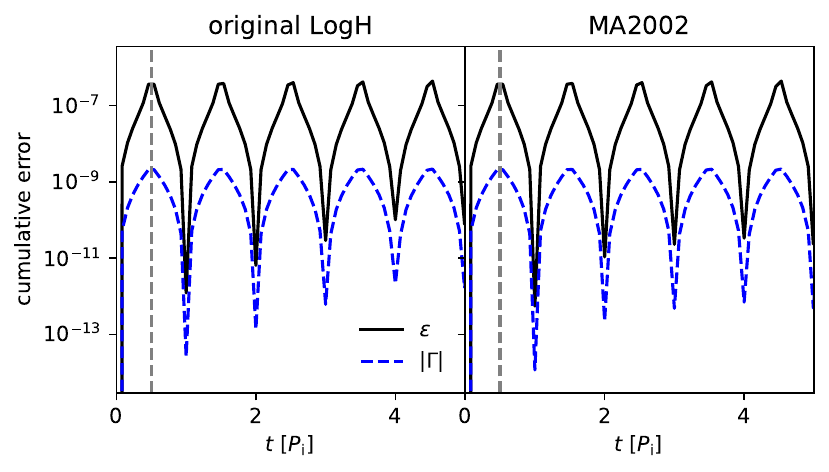}
    \caption{Cumulative absolute integration errors for the same triple, including a weakly perturbed binary, as shown in Figure~\ref{fig:compbintr}, using the original LogH and MA2002 methods with $\Delta s_0/256$. The plotting style is similar to Figure~\ref{fig:binge}.}
    \label{fig:comptrge}
\end{figure}

The behaviors of the two methods for triples, including a weakly perturbed binary discussed in Section~\ref{sec:binary-triple}, are similar. 
Figure~\ref{fig:comptrge} compares $\epsilon$ and $\Gamma$ for the triple using both methods with $\Delta s_0/256$.
Both methods yield identical results.

\begin{figure}[ht!]
    \centering
    \includegraphics[width=0.7\linewidth]{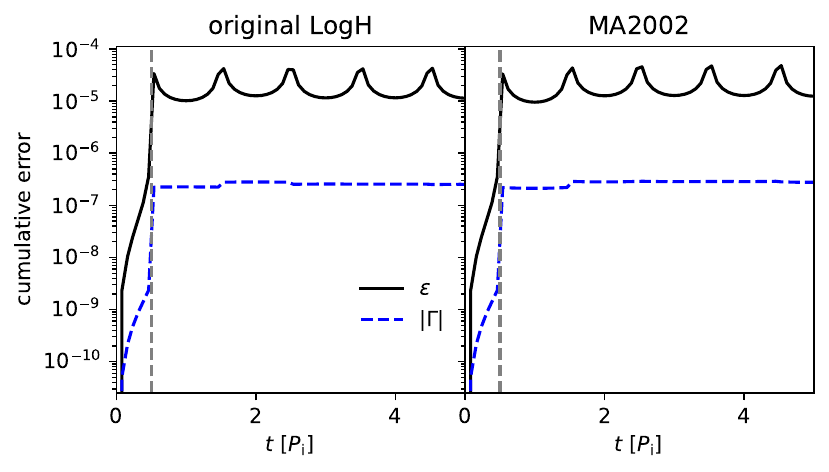}
    \caption{Cumulative absolute integration errors for triples, similar to Figure~\ref{fig:comptrge}, but with the inner binary eccentricity set to $1-e = 1\times 10^{-8}$.}
    \label{fig:comptrgehe}
\end{figure}

Furthermore, we compare the evolution of another triple system by changing $1-\ein$ to $1\times 10^{-8}$, as shown in Figure~\ref{fig:comptrgehe}. 
Unlike the isolated binary case in Figure~\ref{fig:binge}, the errors of both $\epsilon$ and $\Gamma$ increase significantly after the first periapsis and do not return. 
Both methods again yield identical results.
The round-off error differences observed in the binary case are diminished by the approximation error in the two triple cases.
Therefore, in this work, we only present the MA2002 results, which sufficiently represent the behavior of both methods for triples. 


%


\bibliography{paper}{}
\bibliographystyle{aasjournal}


\end{CJK*}
\end{document}